\documentclass[12pt]{spieman}  
\usepackage{amsmath,amsfonts,amssymb}
\usepackage{graphicx}
\usepackage{setspace}
\usepackage{tocloft}
\usepackage{array}
\usepackage{multirow}
\usepackage{upgreek}

\newcommand{\PreserveBackslash}[1]{\let\temp=\\#1\let\\=\temp}
\newcolumntype{C}[1]{>{\PreserveBackslash\centering}p{#1}}
\newcolumntype{L}[1]{>{\PreserveBackslash\raggedright}p{#1}}
\newcommand{\micron}[0]{$~\upmu$m }

\newcommand{\new}[1]{{#1}}

\title{Optical Verification Experiments of Sub-scale Starshades}

\author[a,*]{Anthony Harness}
\author[b]{Stuart Shaklan}
\author[b]{Phillip Willems}
\author[a,c]{N. Jeremy Kasdin}
\author[b]{K. Balasubramanian}
\author[b]{Philip Dumont}
\author[b]{Victor White}
\author[b]{Karl Yee}
\author[b]{Rich Muller}
\author[a]{Michael Galvin}
\affil[a]{Princeton University, Mechanical \& Aerospace Engineering Department, Princeton, New Jersey}
\affil[b]{Jet Propulsion Laboratory, California Institute of Technology, Pasadena, California}
\affil[c]{University of San Francisco, College of Arts and Sciences, San Francisco, California}

\cftpagenumbersoff{figure}
\cftpagenumbersoff{table}
\begin{document}
\maketitle

\begin{abstract}
Starshades are a leading technology to enable the detection and spectroscopic characterization of Earth-like exoplanets. In this paper we report on optical experiments of sub-scale starshades that advance critical starlight suppression technologies in preparation for the next generation of space telescopes. These experiments were conducted at the Princeton starshade testbed, an 80 m long enclosure testing 1/1000$^\text{th}$ scale starshades at a flight-like Fresnel number. We demonstrate 10$^{-10}$ contrast at the starshade's geometric inner working angle across 10\% of the visible spectrum, with an average contrast at the inner working angle of $2.0\times10^{-10}$ and contrast floor of $2\times10^{-11}$. In addition to these high contrast demonstrations, we validate diffraction models to better than 35\% accuracy through tests of intentionally flawed starshades. Overall, this suite of experiments reveals a deviation from scalar diffraction theory due to light propagating through narrow gaps between the starshade petals. We provide a model that accurately captures this effect at contrast levels below $10^{-10}$. The results of these experiments demonstrate that there are no optical impediments to building a starshade that provides sufficient contrast to detect Earth-like exoplanets. This work also sets an upper limit on the effect of unknowns in the diffraction model used to predict starshade performance and set tolerances on the starshade manufacture.
\end{abstract}

\keywords{Starshades, optical model validation, high contrast imaging, exoplanet detection, vector diffraction}

{\noindent \footnotesize\textbf{*}Send email correspondence to: Anthony Harness, \href{mailto:aharness@princeton.edu}{aharness@princeton.edu} }

\begin{spacing}{2}   

\section{Introduction}
\label{sec:intro}

Starshades have the potential to discover and characterize the atmospheres of Earth-like exoplanets in the habitable zone of nearby stars\cite{Cash_2006, WFIRST_SRM,  HabEx}. Their ability to achieve high contrast while maintaining high optical throughput and broad wavelength coverage make them the most promising technology to produce the first spectrum of an exo-Earth atmosphere\cite{Theia, Turnbull_2012}. In recent years there have been significant technological advances that demonstrate the feasibility of building and deploying a starshade\cite{S5_Plan, ExEP_Tech, S5_JATIS}, which have led to a significant increase in the community's interest in a future starshade mission. There is interest in a starshade to rendezvous with NASA's next flagship mission,  the Nancy Grace Roman Space Telescope\cite{WFIRST_SRM}, and a starshade is baselined for the proposed flagship mission, the Habitable Exoplanet Observatory (HabEx)\cite{HabEx}.

Though significant progress has been made, starshades are still an unproven technology, particularly with respect to their optical performance. The distributed architecture's size (10's of meters diameter over 10,000's of kilometers) and sensitivity (10$^{10}$ relative change in intensity) are unprecedented---when constructed, the starshade will be the largest visible-light optic ever made. \new{Consequently, we need a reliable means to experimentally validate the design tools that rely on an accurate prediction of the optical performance. The state of the art optical models\cite{Cash_2011, Vanderbei_2007, Cady_2012, Harness_2018} operate under the assumptions of: a scalar theory of diffraction; scale invariance of the Fresnel approximation; and a scalar application of Babinet's principle. The work presented here focuses on demonstrating the validity of the first assumption.}

The diffraction equations employed in the starshade context are invariant with Fresnel number, so while it is impossible to test a full-scale starshade on the ground due to its size, we can validate the optical models with sub-scale experiments if conducted at a flight-like Fresnel number. Previous experiments in Refs.~\citenum{Leviton_2007, Samuele_2009, Cady_2009, Samuele_2010, Sirbu_2011, Glassman_2014, Glassman_2015} were conducted at Fresnel numbers much larger than that expected in flight; the experiment in Ref.~\citenum{Harness_2017} was done at a flight-like Fresnel number, but its contrast was limited by the atmosphere. With the Princeton starshade testbed presented here, we are for the first time able to achieve 10$^{-10}$ contrast with a starshade at a flight-like Fresnel number.

The Starshade to TRL 5 (S5) project was established by the NASA Exoplanet Exploration Program to advance starshade technology to TRL 5 in a time frame compatible with a starshade rendezvous with the Roman mission\cite{S5_Plan,S5_JATIS}. The work presented here was conducted under S5 to advance the seminal optical technology of starshades---starlight suppression. We group the experiments into two categories, ``optical verification'' and ``model validation'', which reflect the two milestones set as the criteria needed to reach TRL 5\cite{S5_Plan}. Optical verification experiments show we can design an apodization function, which specifies the starshade's shape, that provides sufficient contrast to achieve our stated science goals. These experiments validate the fundamental operation of the starshade and demonstrate that the aforementioned assumptions are valid. Model validation experiments show that the optical models correctly capture the performance sensitivity to perturbations in the starshade shape and set an upper limit to the model uncertainty used in design tools to derive the starshade shape error budget and tolerances for future missions.

To briefly summarize our main results, we have demonstrated 10$^{-10}$ contrast at the geometric inner working angle (IWA) of a starshade with a flight-like Fresnel number across a 10\% bandpass while reaching a contrast floor of $2\times10^{-11}$ beyond the starshade tips. From this we conclude that we can predict the nominal starshade performance to at least 10$^{-10}$ contrast. The main limitation to the contrast at the laboratory scale comes from a polarization-dependent effect of light propagating through the narrow gap between starshade petals. We provide an explanation of this effect in Sec.~\ref{sec:thick_screen}. \new{The model validation experiments demonstrate better than 35\% (with an average of 20\%)
model accuracy for a number of different shape perturbations and at multiple wavelengths. This result means we must only carry a contrast margin of 1.35$\times$ in the design error budget. The results from these experiments build confidence in our ability to successfully design a starshade that will provide the contrast needed to detect Earth-like exoplanets.}

In Sec.~\ref{sec:design} we describe the layout and individual components of the starshade testbed and outline the experiments performed. Section~\ref{sec:verification_experiments} presents the results of the optical verification experiments, Sec.~\ref{sec:validation_experiments} presents the results of the model validation experiments, and we discuss the implications of these results in Sec.~\ref{sec:discussion}. We summarize and conclude in Sec.~\ref{sec:conclusion}. Additional details and results from the optical verification experiments can be found in the S5 milestone final reports\cite{Mile1a, Mile1b}, which have been accepted by the Exoplanet Exploration Program Technical Advisory Committee\cite{TAC}.

\section{Experiment Design}
\label{sec:design}
\renewcommand*{\theHsection}{chX.\the\value{section}}

The experiments presented here were conducted at the Princeton starshade testbed, a dedicated facility in the Frick chemistry building on the Princeton campus. The testbed was designed to replicate the flight configuration at 1/1000$^\text{th}$ scale as closely as is possible given the differences in size and environment. The scale of the experiment is ultimately limited by the longest separation available in an indoor facility on campus. We use the Starshade Rendezvous Probe Mission\cite{WFIRST_SRM} (SRM) and HabEx mission\cite{HabEx} as the reference flight configurations; parameters for the flight and laboratory configurations are presented in Table~\ref{tab:ss_params}. The table lists the operational range of the Fresnel number, $N$, defined as
\begin{equation}
    N=\dfrac{R^2}{\lambda Z_\text{eff}} \,,
    \label{eq:fresnel_number}
\end{equation}
where $R$ is the starshade radius, $\lambda$ is the wavelength of light, and $Z_\text{eff}$ is \new{the effective starshade-telescope separation:
\begin{equation}
    Z_\text{eff} = \frac{Z_\text{src}Z_\text{tel}}{Z_\text{src} + Z_\text{tel}} \,,
    \label{eq:z_eff}
\end{equation}
where $Z_\text{tel}$ is the distance between telescope and starshade and $Z_\text{src}$ is the distance between starshade and light source. $Z_\text{eff}$ accounts for the finite distance to the diverging beam light source in the laboratory configuration and is derived in Eq.~(\ref{eq:fresN_part1}). For the laboratory experiments, $Z_\text{eff}=17.7$ m; for the flight configuration, the source (target star) is effectively infinitely far away, making $Z_\text{eff}\approx Z_\text{tel}$.

In this work we use the geometric IWA (~$= R/Z_\text{tel}$) as the point of reference, though future design studies should follow Ref.~\citenum{Arenberg_2007} and set requirements relative to the effective IWA, which accounts for the width of the telescope's point spread function (PSF).}
\begin{table}[ht]
\caption{Optical parameters for the laboratory experiments and the SRM and HabEx starshade architectures. The range of Fresnel numbers is set by the wavelength range. $^\dagger$For apodization design C12/C16 in Table~\ref{tab:apod_design}.}
\label{tab:ss_params}
\begin{center}
\begin{tabular}{ l c c c }
\hline
\rule[-1ex]{0pt}{3.5ex} & Laboratory & SRM & HabEx \\
\hline\hline
\rule[-1ex]{0pt}{3.5ex} Telescope diameter $(D)$ & 5.0 mm & 2.37 m & 4.0 m\\
\rule[-1ex]{0pt}{3.5ex} Starshade diameter $(2R)$ & 25.06 mm$^\dagger$ & 26 m & 52 m\\
\rule[-1ex]{0pt}{3.5ex} Wavelength $(\lambda)$ range & 641 - 725 nm$^\dagger$ & 616 - 800 nm & 300 - 1000 nm\\
\rule[-1ex]{0pt}{3.5ex} Telescope - starshade sep.  $(Z_\text{tel})$ & 50.0 m & 26,000 km & 76,600 km\\
\rule[-1ex]{0pt}{3.5ex} Source - starshade sep. $(Z_\text{src})$ & 27.45 m & $>3$ parsec & $>3$ parsec\\
\rule[-1ex]{0pt}{3.5ex} {\bf Fresnel number} $(N)$ & {\bf12.2 $-$ 13.8}$^\dagger$ & {\bf 8.1 $-$ 10.5} & {\bf8.8 $-$ 29}\\
\hline
\end{tabular}
\end{center}
\end{table}

For the range of $N$ under study, the Fresnel approximation of diffraction is sufficiently accurate\cite{Harness_2018} to compute the diffraction pattern to contrast levels better than $10^{-10}$, an assertion borne out by the successful demonstration of a dark shadow in these experiments. \new{We use the Fresnel-Kirchhoff\cite{Born_Wolf} diffraction equation to describe the diffraction in both the laboratory and flight configurations. We invoke the standard paraxial and Fresnel approximations and assume circular symmetry (with radial coordinate at the starshade $r$). The electric field incident on the starshade $(U_0)$ is due to a spherical wave of amplitude $u_0$ emanating at a distance $Z_\text{src}$:
\begin{equation}
    U_0\left(r\right) = \frac{u_0}{Z_\text{src}}\exp{\left\{\frac{i\pi r^2}{\lambda Z_\text{src}}\right\}} \,.
\end{equation}

We assume the starshade's shape is an approximation of a smooth radial apodization function, $A\left(r\right)$, a valid approximation given a sufficient number of petals\cite{Vanderbei_2007}. The Fresnel-Kirchhoff diffraction integral to compute the on-axis electric field $U$ at the telescope pupil plane (dropping the leading phase factor) is given by
\begin{eqnarray}
 	U &  = & \frac{2\pi}{i\lambda Z_\text{tel}} \int_0^{R} A\left(r\right)U_0\left(r\right)\exp{\left\{\frac{i\pi r^2}{\lambda Z_\text{tel}}\right\}}r\, dr \nonumber \\
    &  = & \frac{2\pi u_0}{i\lambda Z_\text{tel}Z_\text{src}} \int_0^{R} A\left(r\right)\exp{\left\{\frac{i\pi r^2}{\lambda Z_\text{src}}\right\}}\exp{\left\{\frac{i\pi r^2}{\lambda Z_\text{tel}}\right\}}r\, dr \nonumber \\
 &  = & \frac{2\pi u_0}{i\lambda Z_\text{tel}Z_\text{src}} \int_0^{R} A\left(r\right)\exp{\left\{\frac{i\pi r^2}{\lambda}\left(\frac{1}{Z_\text{src}} + \frac{1}{Z_\text{tel}}\right)\right\}}r\, dr \,.
    \label{eq:fresN_part1}
\end{eqnarray}

The distance terms in the exponential can be combined into an effective separation parameter,  $Z_\text{eff}$, given by Eq.~(\ref{eq:z_eff}). We set the amplitude of the incident wave to be unity at the starshade, to give $u_0 = Z_\text{src}$. Equation~(\ref{eq:fresN_part1}) can be rewritten as
\begin{eqnarray}
 	U &  = & \frac{2\pi}{i\lambda Z_\text{tel}} \int_0^{R} A\left(r\right)\exp{\left\{\frac{i\pi r^2}{\lambda Z_\text{eff}}\right\}}r\, dr \nonumber \\
     & = & \frac{\pi}{i}\frac{Z_\text{eff}}{Z_\text{tel}} \int_0^{N} A(n)\exp{\left\{i\pi n\right\}}\, dn\,,
    \label{eq:fresN}
\end{eqnarray}
with dimensionless quantity $n~=~r^2/\lambda Z_\text{eff}$ spanning the range of Fresnel numbers up to $N$.

Written in the dimensionless form of Eq.~(\ref{eq:fresN}) and neglecting the constant amplitude scale factor $\approx 1$, the integral is independent of $R$, $\lambda$, and $Z_\text{eff}$, and depends solely on the Fresnel number. This enables experimental validation via more practical sub-scale experiments: by showing in the laboratory that high contrast is achieved at flight-like values of $N$, we show that the full-scale, properly shaped starshade will also produce the same level of contrast. This statement is true under the assumption that scalar diffraction theory holds at both scales. The experiments described here show that scalar diffraction theory almost holds at small scales, as long as it is corrected with an additive non-scalar component at the edges.  Extending the same modeling approach to flight scale shows that the additive terms makes a negligible contribution (see Appendix~\ref{apx:skin_scaling}), validating the extension of the scalar diffraction theory result to flight scale.

Due to size constraints of the testbed, the laboratory Fresnel number does not extend as low\footnote{The lowest Fresnel number tested is $N=11.5$ in Sec.~\ref{sec:varying_fresnel}.} as those in the full scale configurations, but still falls within the range of Fresnel numbers set by the span in wavelength. Additionally, the Fresnel number ($N$) quoted in Table~\ref{tab:ss_params} is in reference to the maximum radius of the starshade, but the diffraction integral starts at the base of the petals with $n\sim5$ and integrates to $N$. This means our experiments will demonstrate that we accurately capture the behavior of the apodization function over the range of Fresnel numbers seen in the flight configurations. Testing at $N<15$ has been determined to be sufficiently adequate for the purposes of model validation\cite{S5_Plan}.}

Altogether, by utilizing a 80 m long facility, we are able to test at a flight-like Fresnel number a starshade that is large enough to be accurately manufactured with existing technology.

\subsection{Summary of experiments}
The experiments presented here can be sorted into two loose categories that track the two S5 milestones\cite{S5_Plan} that this work is tasked to complete: optical verification\cite{Mile1a, Mile1b} (presented in Sec.~\ref{sec:verification_experiments}) and model validation (presented in Sec.~\ref{sec:validation_experiments}). Table~\ref{tab:experiment_summary} provides a summary of six experiments and the name (production number) of the starshade tested. Improvements to the testbed were made between various experiments, the most dramatic of which was the installation of a linear polarizer after the fiber launcher (see Sec.~\ref{sec:light_source}) and a linear polarizer as analyzer on a rotation stage in front of the camera. This change occurred after completing experiments 1, 2, and 6 of Table~\ref{tab:experiment_summary}.

\begin{table}[ht]
\caption{Summary of experiments and production number of starshades tested (see Table~\ref{tab:masks} for details on specific starshades). Those with OV in the Goal column are optical verification experiments and are presented in Sec.~\ref{sec:verification_experiments}. Those with MV in the Goal column are model validation experiments and are presented in Sec.~\ref{sec:validation_experiments}.
\label{tab:experiment_summary}
}
\begin{center}
\begin{tabular}{c L{2.5cm} c L{7.5cm} c L{1.5cm}}
\hline
\rule[-1ex]{0pt}{3.5ex} \# & Experiment & Section & Brief description & Goal & Starshade\\
\hline\hline
\rule[-1ex]{0pt}{4.5ex}1 & Monochromatic contrast & \ref{sec:mono} & Demonstrate best possible contrast at a single wavelength.  & \multirow{2}{*}{OV} & \multirow{2}{*}{DW17}\\
\rule[-1ex]{0pt}{4.5ex}2 & Broadband contrast & \ref{sec:broadband} & Demonstrate $<$ 10$^{-10}$ contrast at the inner working angle across a 10\% bandpass. & \multirow{2}{*}{OV} &\multirow{2}{*}{DW21}\\
\rule[-1ex]{0pt}{4.5ex}3 & Exposed \hspace{6pt} petal tips & \ref{sec:exposed_tips} & Demonstrate high contrast with a starshade with realistic petal tips. & \multirow{2}{*}{OV} & \multirow{2}{*}{M12P3}\\
\rule[-1ex]{0pt}{4.5ex}4 & Shape perturbations & \ref{sec:perturbation_experiments} & Validate optical models against starshades with perturbations built into their shape. & \multirow{2}{*}{MV} & M12P2~/ M12P3\\
\rule[-1ex]{0pt}{4.5ex}5 & Polarization study & \ref{sec:polarization} & Improve non-scalar diffraction model by studying polarization-dependent effects. & \multirow{2}{*}{MV} & DW9~/ DW21\\
\rule[-1ex]{0pt}{4.5ex}6 & Varying Fresnel \# & \ref{sec:varying_fresnel} & Validate Fresnel number dependence of model by testing at lower Fresnel number. & \multirow{2}{*}{MV} & \multirow{2}{*}{DW17}\\
\hline
\end{tabular}
\end{center}
\end{table}

\subsection{Testbed configuration}
\begin{figure}
\begin{center}
\begin{tabular}{c}
  \includegraphics[width=\linewidth]{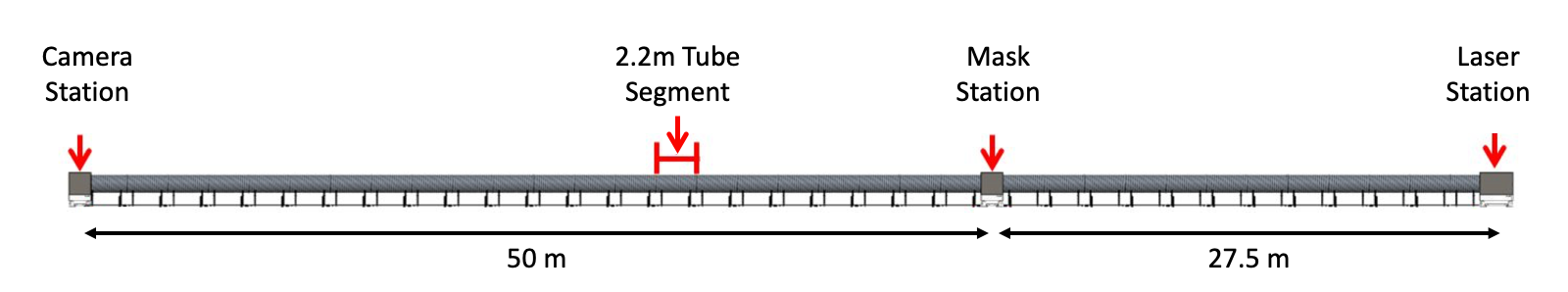}
\end{tabular}
\end{center}
\caption
{ \label{fig:testbed}
Layout of testbed showing distances between camera, starshade mask, and laser stations.}
\end{figure}

The design of the experiment is simple: image a light source from within the deep shadow created by a starshade and measure the efficiency with which the starshade suppresses the on-axis light. The testbed (shown schematically in Fig.~\ref{fig:testbed}) consists of three stations containing a laser, starshade, and camera. The main driver in the testbed design was to maximize the starshade size while maintaining a flight-like Fresnel number, which translates to maximizing the separation. The testbed design is set by the longest, straight-line facility to be found on campus, which gives a total testbed length of 80 m. Since a fraction of the length is needed for propagation of the diverging beam, the effective separation between starshade and telescope is 17.7 m, which sets the starshade size to 25 mm diameter.

The beam line is contained in 1 m diameter steel tubing (not a vacuum) to seal the testbed from stray light and dust and to help stabilize the atmosphere. The tube is wrapped in fiberglass insulation to minimize the effect of external thermal changes. All equipment is built to be remotely operated to minimize how often the testbed is opened, which generates atmospheric turbulence and stirs up dust. Additional details of the testbed design can be found in Refs.~\citenum{Galvin_2016,Kim_2016,Mile1a}.

\subsection{Light source}
\label{sec:light_source}
\begin{figure}
\begin{center}
\begin{tabular}{c}
  \includegraphics[width=\linewidth]{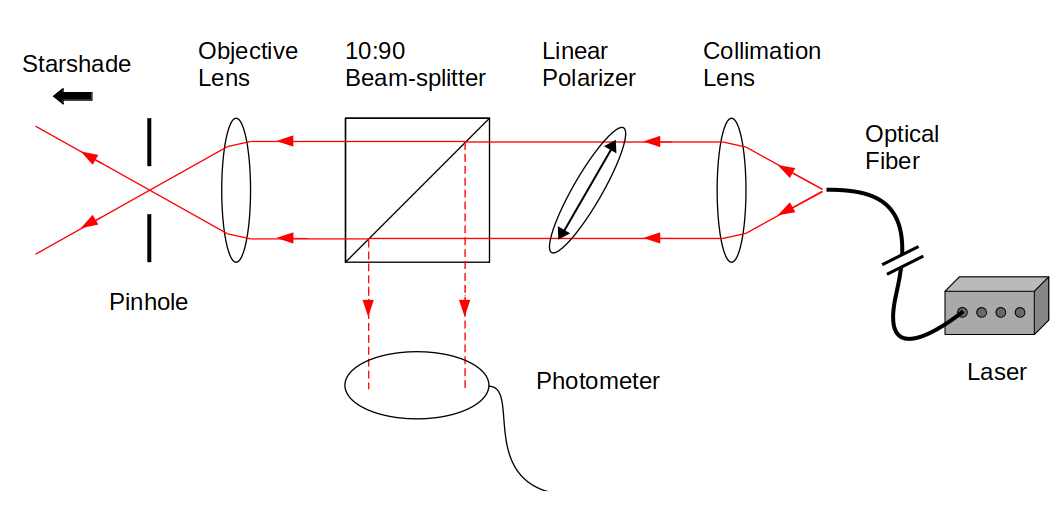}
\end{tabular}
\end{center}
\caption
{ \label{fig:beamsplitter}
\new{Cartoon diagram of laser launching system. The optical fiber enters on the right from the laser outside the testbed. Light is launched from the fiber, collimated, and passes through a linear polarizer before reaching a beam-splitter. 10\% of the light is reflected to a photometer to record the throughput. The other 90\% continues to an objective lens which sends a diverging beam to the starshade. A pinhole at the focus spatially filters high-order aberrations.}}
\end{figure}

The light source serving as the artificial star is a multi-channel laser diode operating at: 405 nm, 638 nm, 641 nm, 660 nm, 699 nm, and 725 nm. The 405 nm light is outside the starshade's operating bandpass and is used for alignment. The laser is located outside the enclosure and fed in via a polarization-maintaining single-mode fiber optic. The polarization out of the fiber depends on external environmental conditions, which vary with time. The fiber terminates with a collimator and the output gaussian beam is focused by an objective lens through a pinhole to spatially filter high-order aberrations. Experiments \# 1, 2, and 6 of Table~\ref{tab:experiment_summary} were done using the resultant polarization out of the fiber. In the rest of the experiments, a linear polarizer is placed between the collimator and objective lens and is fixed horizontally (in the image and lab frames). As the state of polarization out of the fiber varies with external conditions, the power transmitted through the polarizer (and ultimately incident on the starshade) varies with time. To account for this, a beam-splitter after the polarizer sends a fraction of the light to a photometer, which records the transmitted power during observations and allows for the contrast calibration to be adjusted accordingly. A cartoon diagram of the laser launching system is shown in Fig.~\ref{fig:beamsplitter}.

\subsection{Starshade masks}
\begin{figure}
\begin{center}
\begin{tabular}{c}
  \includegraphics[width=\linewidth]{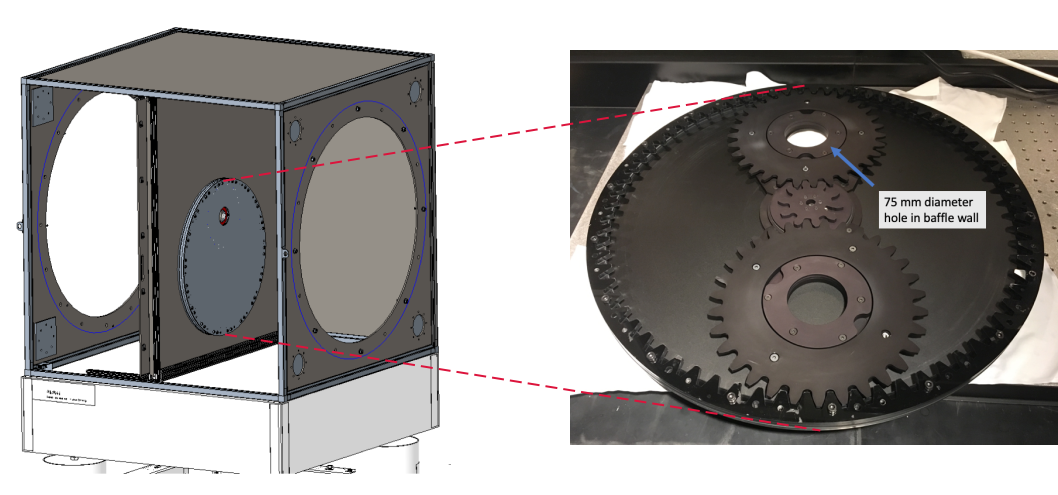}
\end{tabular}
\end{center}
\caption
{ \label{fig:mask_changer}
\new{Left: starshade station located 50 m from the camera. A wall blocks all light expect that passing through the starshade. Right: mask holder with a motorized planetary gear that holds the starshade and calibration masks.}}
\end{figure}

The starshades tested are roughly 25 mm in diameter and are etched into a 100 mm silicon wafer. They are positioned in the middle testbed station and are held by a mask changer (shown in Fig.~\ref{fig:mask_changer}) with a motorized planetary gear that allows us to switch between starshade and calibration masks and to image the mask at different rotation angles.

\subsubsection{Design}
\begin{figure}
\begin{center}
\begin{tabular}{c}
  \includegraphics[width=\textwidth]{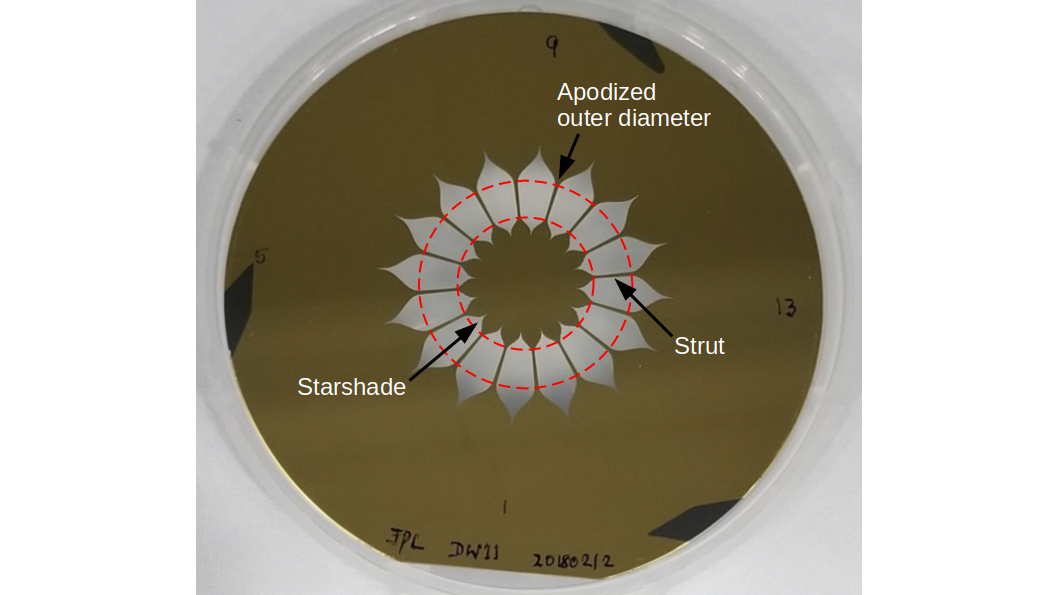}
\end{tabular}
\end{center}
\caption{
\label{fig:mask_image}
Starshade pattern etched into a SOI wafer, manufactured at Microdevices Lab at JPL. Interior to the inner red circle is the inner starshade representing a free floating occulter. The inner starshade is supported in the wafer by radial struts. The outer red circle marks the start of the apodization function of the outer diameter.}
\end{figure}

The starshade mask (shown in Fig.~\ref{fig:mask_image}) consists of an inner starshade, representative of a free floating occulter, that is supported in a silicon wafer via radial struts. The outer diameter of the support wafer is also apodized to minimize the diffraction \new{that would occur from the truncation of the beam by the outer diameter\cite{Cady_2009}. This design results in the starshade mask consisting of $N_p$ ( = number of petals) transmission regions bounded by the petals of the inner starshade, the radial struts, and petals of the outer diameter.}

Both apodization profiles (inner and outer) are designed independent of each other using the numerical optimization scheme outlined in Ref.~\citenum{Vanderbei_2007}. \new{Table~\ref{tab:apod_design} details the designs of a number of apodization functions; the minimum radius $(R_0)$ is the radius at which the inner petals start and the maximum radius $(R)$ is the radius at which the struts start (where the tips of a free floating occulter would be). Design A of Table~\ref{tab:apod_design} is an earlier design with smaller gaps and a larger maximum radius. Design B was specifically designed for these experiments. We impose a constraint on the radius of the inner starshade to have Fresnel number $<15$ and constrain the gaps between the starshade petals to have widths $>$~16\micron to minimize non-scalar diffraction. We found this to be the largest gap width that provides a valid solution to the optimization problem. Design C16 is the same apodization profile as Design B, but is made 3\% larger to shift the operating bandpass to cover the laser's available wavelength channels. Design C12 is the same as Design C16, but with 12 petals instead of 16, which was done to minimize the number of inner gaps between petals, which serve as sources of non-scalar diffraction.}

\begin{table}[ht]
\caption{Design of apodization functions including the number of petals, operating bandpass, minimum radius (start of the petals), maximum radius (start of the struts), and gap width between petals.}
\label{tab:apod_design}
\begin{center}
\begin{tabular}{ c c c c c c }
\hline
\rule[-1ex]{0pt}{3.5ex}\multirow{2}{*}{Design} & Number & Operating & Minimum & Maximum & \multirow{2}{*}{Gap Width}\\
\rule[-1ex]{0pt}{3.5ex} & of Petals & Bandpass & Radius $(R_0)$ & Radius $(R)$ & \\
\hline\hline
\rule[-1ex]{0pt}{3.5ex}A & 16 & 600 nm - 670 nm & 8.41 mm & 17.35 mm & 7.5\micron \\
\rule[-1ex]{0pt}{3.5ex}B & 16 &  600 nm - 690 nm & 8.02 mm & 12.15 mm & 16.2\micron \\
\rule[-1ex]{0pt}{3.5ex}C16 & 16 &  640 nm - 730 nm & 8.26 mm & 12.53 mm & 16.2\micron \\
\rule[-1ex]{0pt}{3.5ex}C12 & 12 &  640 nm - 730 nm & 8.26 mm & 12.53 mm & 21.6\micron \\
\hline
\end{tabular}
\end{center}
\end{table}

\new{After a solution to the optimization problem is found, the apodization profile is multiplied by 0.9 to provide width to the radial struts and is then petalized to become the starshade shape shown in Fig.~\ref{fig:mask_image}. Since the radial struts consist of a constant multiplication applied to the apodization profile, they do not diffract into the shadow.}

\subsubsection{Manufacturing}
The starshade pattern is etched into the device layer of a silicon-on-insulator (SOI) 100 mm wafer via a deep reactive ion etching process. The allowed tolerances on the shape are very small, $\sim$~100~nm, which is achievable with a direct write electron beam lithography process. The SOI device layer is made as thin as is practical (1\micron - 7~$\upmu$m) to minimize non-scalar diffraction as light propagates past the optical edge. \new{The 350\micron thick support wafer is etched from the backside to recess it 50\micron from the device layer's optical edge. The final step in the process is to coat the top of the device layer with a thin layer of metal to maintain opacity. Either 0.4\micron of gold or 0.25\micron of aluminum is used, both of which have thicknesses more than 50 times greater than their skin depth, so we expect the metal layer to be completely opaque.} Details on the manufactured masks are found in Table~\ref{tab:masks}. We refer the reader to Refs.~\citenum{Bala_2013,Bala_2017} for more details on the manufacturing process.
\begin{table}[ht]
\caption{Descriptions of manufactured masks including the apodization design (detailed in Table~\ref{tab:apod_design}), the thickness of the device layer (optical edge), the thickness and type of metal coating, and the perturbations built into the shape.}
\label{tab:masks}
\begin{center}
\begin{tabular}{ l c c c l }
\hline
\rule[-1ex]{0pt}{3.5ex}\multirow{2}{*}{Name} & Apodization & Edge Thickness & Metal & \multirow{2}{*}{Shape Perturbations}\\
\rule[-1ex]{0pt}{3.5ex} & Design & ($\pm$ 0.5\micron) & Coating & \\
\hline\hline
\rule[-1ex]{0pt}{3.5ex}DW9  & A & 7\micron  & Au - 0.4\micron & None\\
\rule[-1ex]{0pt}{3.5ex}DW17 & B & 2\micron & Au - 0.4\micron & None\\
\rule[-1ex]{0pt}{3.5ex}DW21 & C16 & 3\micron & Au - 0.4\micron & None\\
\rule[-1ex]{0pt}{3.5ex}M12P2 & C12 & 1\micron & Al - 0.25\micron & Displaced edges\\
\rule[-1ex]{0pt}{3.5ex}M12P3 & C12 & 2\micron & Al - 0.25\micron & Sine waves \& exposed tips\\
\hline
\end{tabular}
\end{center}
\end{table}

\subsection{Optics + detector}
The optics system in the camera station has pupil plane and focal plane imaging modes that are toggled by remotely flipping a lens in/out of the optical path. Contrast measurements are made in the focal plane imaging mode with the camera focused to the plane of the light source, simulating an exoplanet observation. We use an f/100 system with a 5 mm diameter aperture, which provides the same number of resolution elements across the starshade as in the flight design. \new{As will be shown in Sec.~\ref{sec:broadband}, the contrast improves as light at the geometric IWA rolls off with the telescope's PSF. A telescope that highly resolves the geometric IWA gets an added boost in contrast. As such, to test in a flight-like configuration, we scale the aperture to conserve the number of resolution elements across the geometric IWA: $n_\text{resolved} = \left(R/Z_\text{tel}\right) / \left(\lambda /D\right)$.}

In pupil imaging mode, we observe the out-of-band diffraction pattern incident on the entrance pupil and use the bright spot of Arago to align the camera with the starshade, precisely what is done in the formation flying scheme to maintain starshade alignment\cite{Bottom_2020, Palacios_2020}.

To perform calibration measurements, a neutral density filter (optical density $>10^{-7}$) is toggled into the optical train by a motorized stage. A linear polarizer on a motorized rotation stage serves as a polarization analyzer. The detector is an Andor iXon Ultra 888 EMCCD with 13\micron pixels. For low noise performance, the detector is operated with its conventional amplifier, i.e., not electron-multiplying, and is liquid cooled down to -90$^\circ$C.

\subsection{Calibrations}
A circular aperture mask is used to calibrate the throughput of the system and convert measurements of the occulted light source to a contrast value (see Appendix~\ref{apx:contrast_definition} for a definition of contrast). The calibration mask is a 50 mm diameter circle etched through a silicon wafer and switches position with the starshade mask via the motorized mask changer. \new{For each set of observations, two sets of images are taken: one with the starshade mask in the beam and one with the calibration mask in the beam. When observing with the calibration mask, a neutral density filter is placed in the optical path. The measured count rate for both sets of images are used in Eq.~(\ref{eq:contrast_final}) of Appendix~\ref{apx:contrast_definition} to calculate contrast. We refer the reader to Ref.~\citenum{Mile1a} for additional details on the calibration process.}

\section{Optical Verification Experiments}
\label{sec:verification_experiments}
The first category of experiments are meant to verify the fundamental concept of a starshade by demonstrating that we can design the starshade's shape to provide the optical performance needed to image exoplanets. These experiments validate most of the assumptions (e.g., binary approximation to a smooth function) made in the equations used to design the apodization function that defines the starshade's shape. Verification is achieved by demonstrating better than 10$^{-10}$ contrast across a wide bandpass with a starshade in a flight-like optical configuration. \new{The results presented in this section represent the completion of Milestones 1A\cite{Mile1a} and 1B\cite{Mile1b} of the S5 Project, which satisfy the first of two main requirements in the Starlight Suppression technology development plan\cite{S5_Plan}.}

\subsection{Monochromatic contrast}
\label{sec:mono}
\begin{figure}[htb]
\begin{center}
\begin{tabular}{c}
  \includegraphics[width=0.8\linewidth]{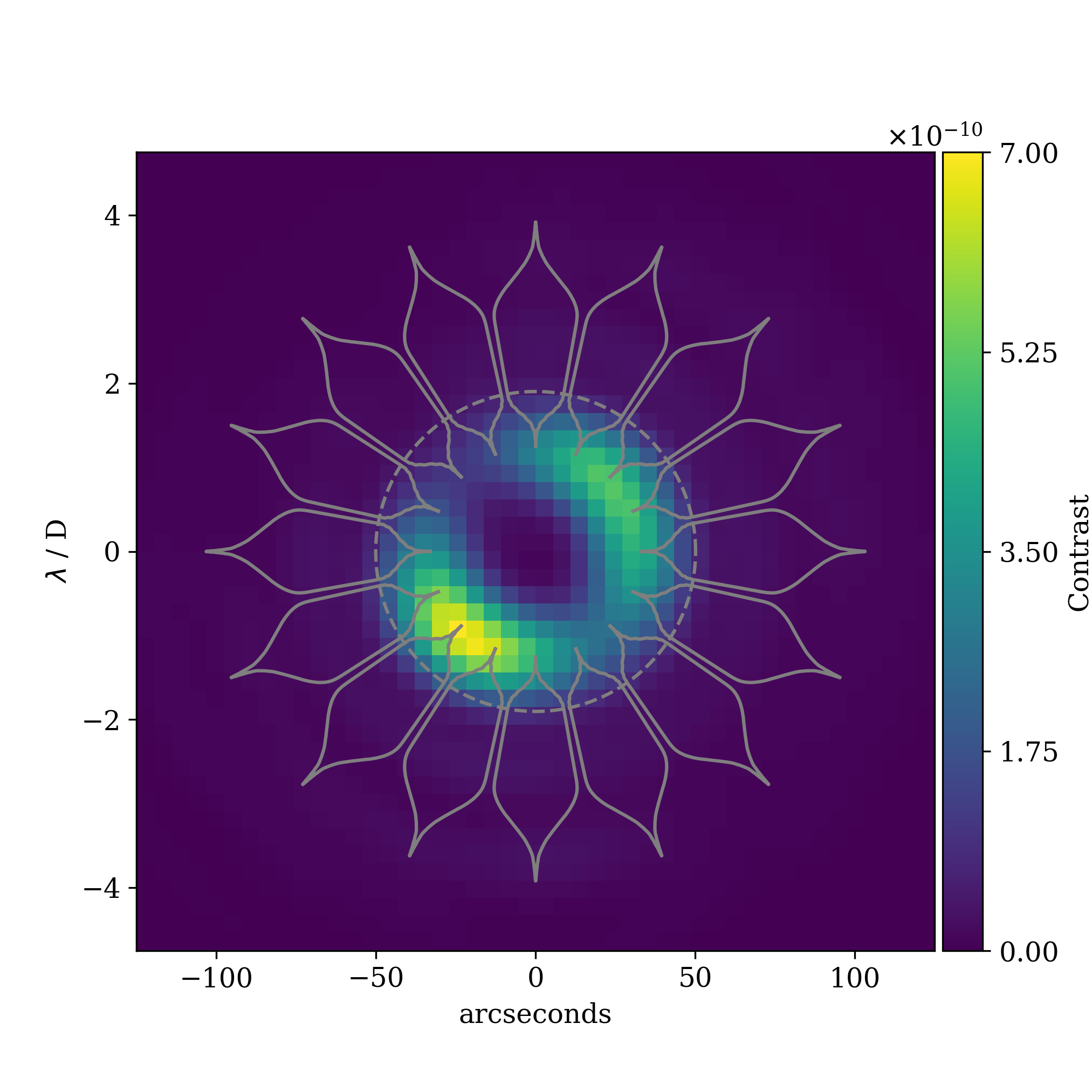}
\end{tabular}
\end{center}
\caption
{ \label{fig:mono_contrast}
Contrast map for monochromatic experiment at $\lambda=638$~nm with mask DW17. The starshade pattern is overlaid and the dashed circle marks the geometric IWA.}
\end{figure}

In this first experiment, we demonstrate the best contrast achieved with the highest quality mask (DW17) at a single wavelength ($\lambda=638$ nm). In the contrast image shown in Fig.~\ref{fig:mono_contrast}, the brightest features are two lobes that are aligned with the polarization vector of the incident light (the intrinsic polarization of the fiber is slightly elliptical at a 40$^\circ$ angle) and which remain fixed as the mask is imaged at different rotation angles. The bright lobes are due to polarization-dependent changes in the electric field as light propagates through the narrow gaps between petals. We call this the ``thick screen effect'' and describe it in detail in Sec.~\ref{sec:thick_screen}. While these lobes are relatively bright at their peak, they are confined to two lobes at the inner gaps between petals and the contrast significantly improves in regions of the image away from the lobes. This means that despite the lobes, 10$^{-10}$ contrast is achieved over a significant fraction of the image at the geometric IWA. This is a key feature of the starshade: any light leaking around the starshade is confined to the edge of the starshade in the image and rolls off with the telescope's PSF at image locations away from the edge.

The primary impact of the lobes is to slightly reduce the image area over which $10^{-10}$ contrast is achieved. The contrast is better than $10^{-10}$ over 44$\%$ of a $\lambda/D$ wide annulus centered at the geometric IWA and quickly rises to 100$\%$ at 1.05$\times$ the IWA\cite{Mile1a}. Figure~\ref{fig:broad_annulus} shows the contrast averaged over a $\lambda /D$ wide annulus as a function of angular separation and simulates the effect of rotating the starshade during the exposure to smear out the diffracted light. The average contrast at the IWA is $1.15\times10^{-10}$ and quickly falls with angular separation to the contrast floor of $2\times10^{-11}$. Figure~\ref{fig:out_of_band} shows that the floor at large angles is set by the non-scalar diffraction lobes and Rayleigh scattering by the atmosphere\cite{Willems_2019}.

\subsubsection{A note on the thick screen effect}
\label{sec:thick_screen}
The bright lobes in Fig.~\ref{fig:mono_contrast} are aligned with the input polarization vector, remain stationary as the starshade rotates, and have a brightness that is an order of magnitude above that predicted by scalar diffraction theory. This was a new discovery that only appeared once high contrast levels were achieved at a flight-like Fresnel number. We have since developed a theory, deemed the thick screen effect\cite{Azpiroz_2003}, which readily explains their origin.

The Fresnel-Kirchhoff diffraction formula, which is used to derive the apodization function represented by the starshade's shape\cite{Vanderbei_2007}, makes the assumption that the electromagnetic field can be represented by a single scalar wave function that satisfies the scalar wave equation\cite{Born_Wolf}, a valid assumption for most optical systems with features larger than the wavelength of light. More specifically, F-K diffraction assumes an infinitely thin, perfectly conducting diffraction screen and that the field in the plane of the screen takes Kirchhoff's boundary conditions, where the field is zero on the screen and is unchanged in the aperture.

The starshades in the lab configuration are small enough that these assumptions begin to break down. The gap between two petals is $\sim$ 20 wavelengths across and the screen (optical edge) is up to half as thick as the gaps are wide, so the gaps resemble waveguides more than thin screens. \new{As light propagates past the thick edge of the starshade mask (through the waveguide), energy is lost due to the finite conductivity of the walls\cite{Jackson}}. The interaction with materials of the edge is polarization-dependent and induces a slight change in the complex field (in both amplitude and phase) and Kirchhoff's boundary conditions are no longer valid -- the propagation of light can no longer be described by scalar diffraction theory alone.

\new{The result of the interaction with the edge is an attenuation of the transmission coefficient through the gap. This can be approximated by a $\sim$ wavelength wide boundary layer\cite{Braunbek} around the starshade's edge that blocks light and changes the effective apodization profile of the petal, negating some of the light suppression. In effect, the width of each petal increases by $\sim\lambda$ on each side.} The strength of the edge effect is roughly the same moving radially outwards; however, the transmission of the apodization profile rapidly increases radially, so the edge effect is strongest in the narrow gaps between petals where it occupies an appreciable fraction of the transmission area. This explains why the lobes are only seen in the innermost regions. The interaction of light with the edge is also dependent on the polarization of the incident light and thus the morphology of the lobes depend on the input polarization state.

Our theory of the edge effect is supported by a newly developed optical model\cite{Harness_2020} and its comparison to experimental results is shown in Sec.~\ref{sec:polarization}. Under this theory, these effects should be negligible at flight scales. The relevant parameter is the width of the thick screen-induced boundary layer relative to the transmission area. The optical edges for a flight starshade will induce roughly the same wavelength wide boundary layer, but the transmission area is 1000$\times$ larger, meaning the contribution from non-scalar diffraction is $10^6\times$ lower and can be considered negligible (see Appendix~\ref{apx:skin_scaling} for a derivation of this argument). Full vector models of flight-scale starshades are beyond the scope of this paper, but will be addressed in the future.

\subsection{Broadband contrast}
\label{sec:broadband}
\begin{figure}[htb]
\begin{center}
\begin{tabular}{c}
  \includegraphics[width=\linewidth]{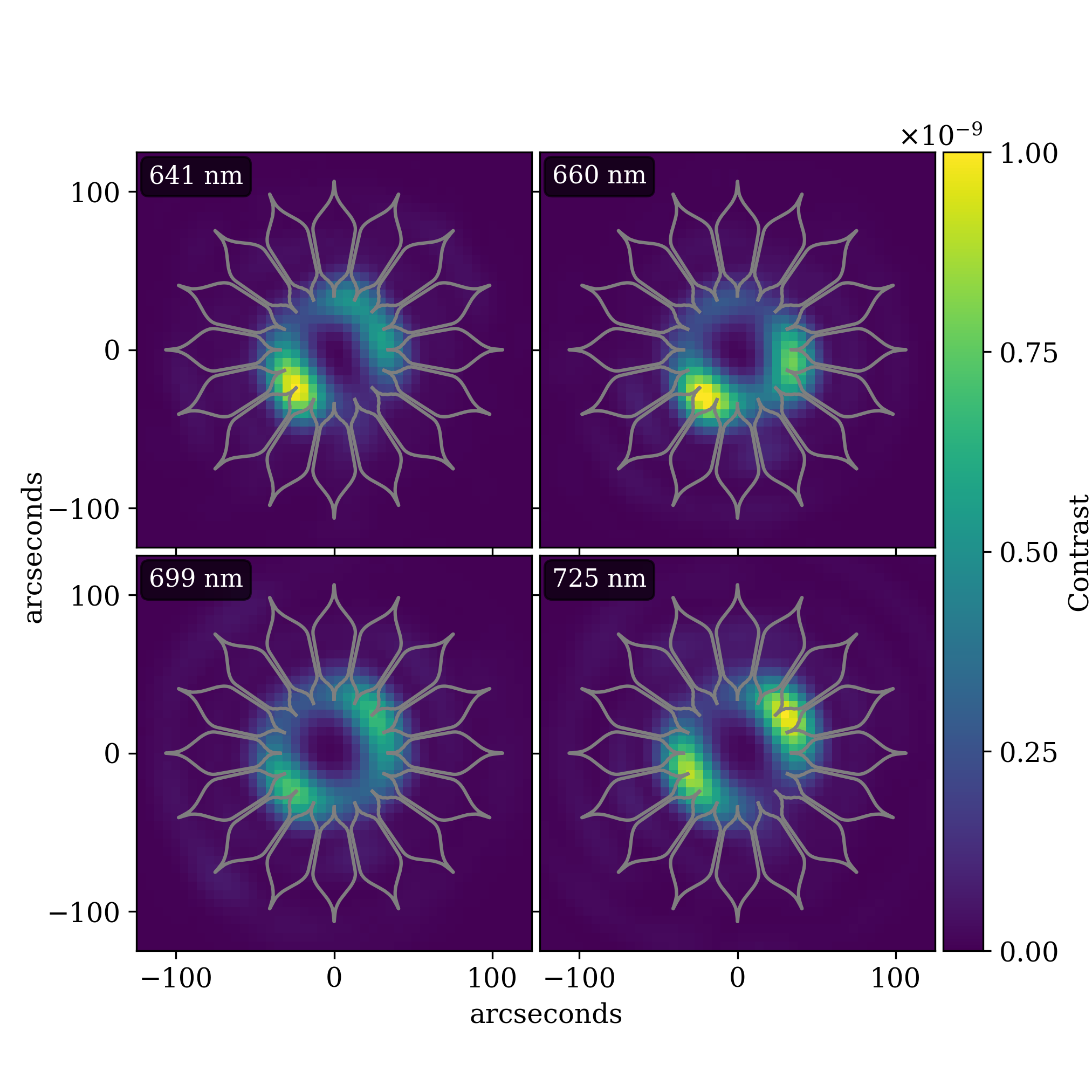}
\end{tabular}
\end{center}
\caption
{ \label{fig:broad_contrast}
Contrast images of mask DW21 at four discrete wavelengths spanning a 10\% bandpass.}
\end{figure}

\begin{figure}[htb]
\begin{center}
\begin{tabular}{c}
  \includegraphics[width=\linewidth]{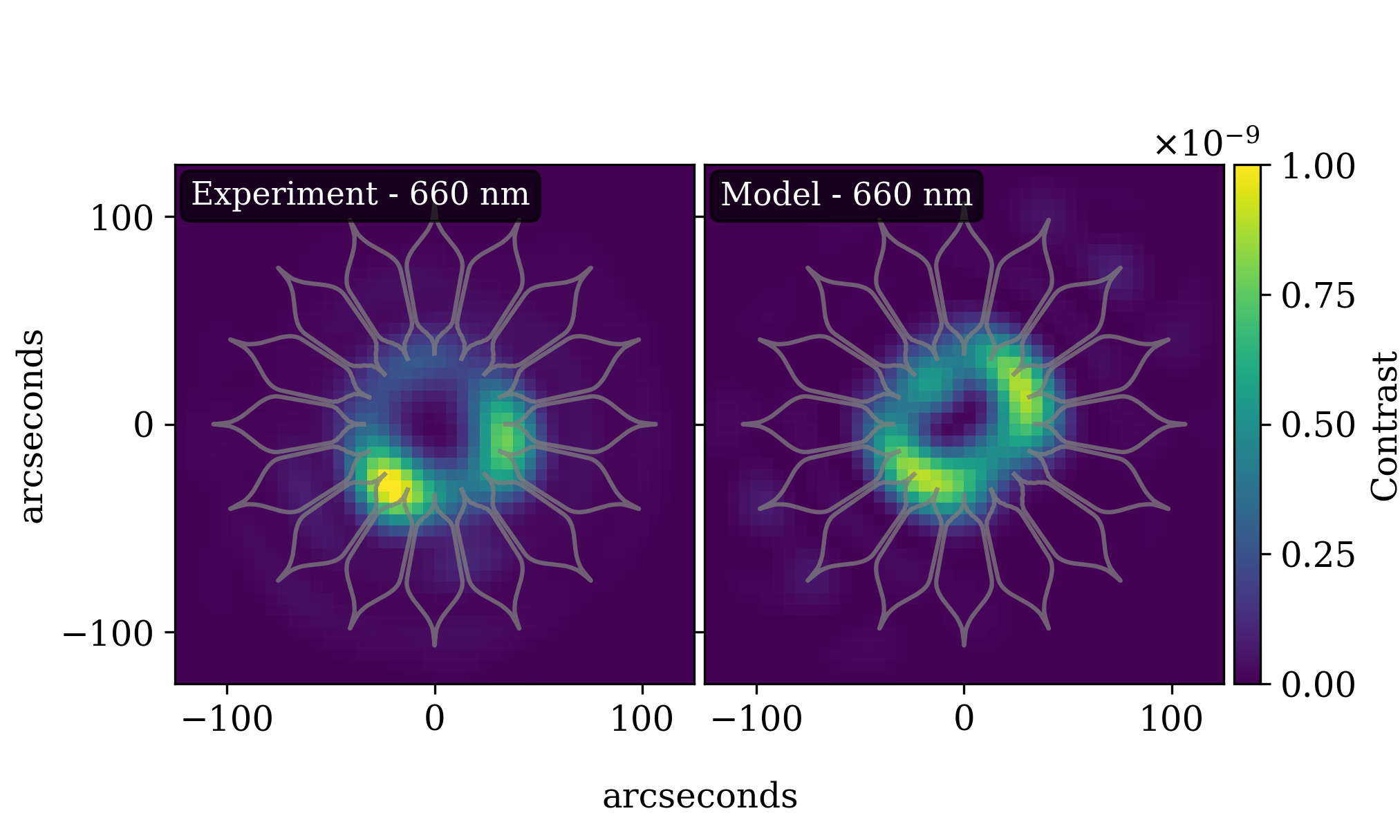}
\end{tabular}
\end{center}
\caption
{ \label{fig:shifted_lobes}
\new{Experiment (left) and model (right) images of mask DW21 at $\lambda=660$~nm. In the model the camera is shifted off-axis by 1 mm, which distorts the polarization lobes, suggesting this can account for the morphology of the lobes seen in the data.}}
\end{figure}

\begin{figure}[htb]
\begin{center}
\begin{tabular}{c}
  \includegraphics[width=0.8\linewidth]{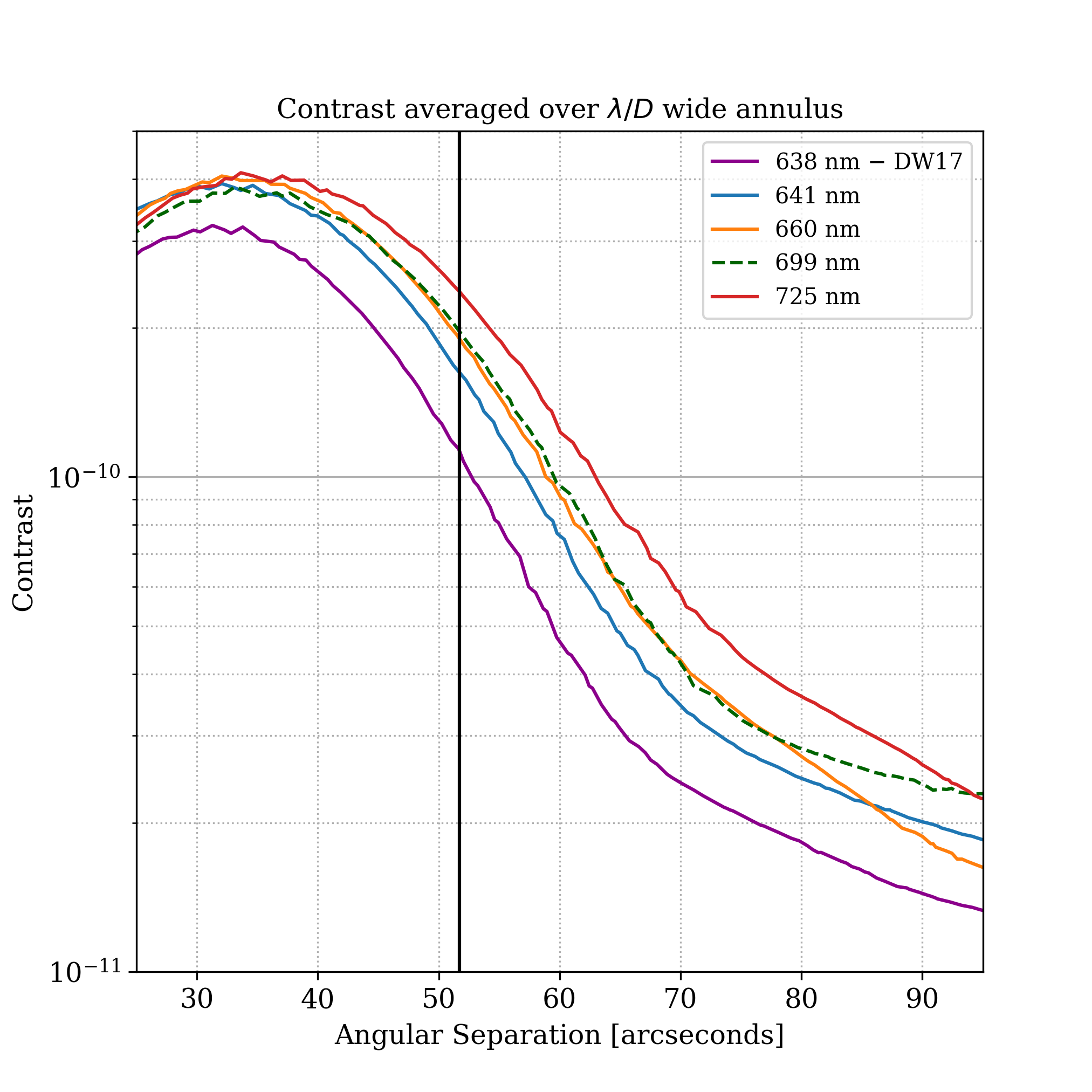}
\end{tabular}
\end{center}
\caption
{ \label{fig:broad_annulus}
Contrast averaged over a $\lambda/D$ wide annulus as a function of angular separation for different wavelength observations of mask DW21. Also included is the monochromatic observation of mask DW17. The solid black line denotes the starshade's geometric IWA.}
\end{figure}

Testing over a wider wavelength range increases the applicability of the experiment to a more flight-like configuration and demonstrates that a starshade can maintain its high contrast performance over a scientifically interesting bandpass. In this experiment we tested mask DW21 at four discrete wavelengths that span a 10\% (85 nm) bandpass. The design of DW21 is identical to, but 3\% larger than, DW17 in order to shift the starshade's operating bandpass to cover the available laser wavelengths. Figure~\ref{fig:broad_contrast} shows again that better than 10$^{-10}$ contrast is achieved at the IWA and that the performance is dominated by the thick screen effect. The contrast is relatively constant across the bandpass, while the peak contrast is higher than in the monochromatic experiment due to DW21 being slightly thicker than DW17, which produces a larger thick screen effect. The morphology of the polarization lobes in the $\lambda=660$~nm data is most likely due to a misalignment between the camera and starshade -- \new{Fig.~\ref{fig:shifted_lobes} shows a model image where the camera is shifted off-axis by 1~mm and the lobes are distorted to one side of the starshade.} Figure~\ref{fig:broad_annulus} shows the contrast averaged over a $\lambda /D$ wide annulus, where the contrast is slightly worse at longer wavelengths as the PSF broadens and more light from the polarization lobes are leaked into the IWA. The average contrast at the IWA across the wavelengths is $2.0\times10^{-10}$. This experiment shows the starshade does not suffer any fundamental degradation in performance by operating across a wider bandpass, as is expected by theory.

\subsection{Exposed tips}
\label{sec:exposed_tips}
\begin{figure}[htb]
\begin{center}
\begin{tabular}{c}
  \includegraphics[width=0.7\linewidth]{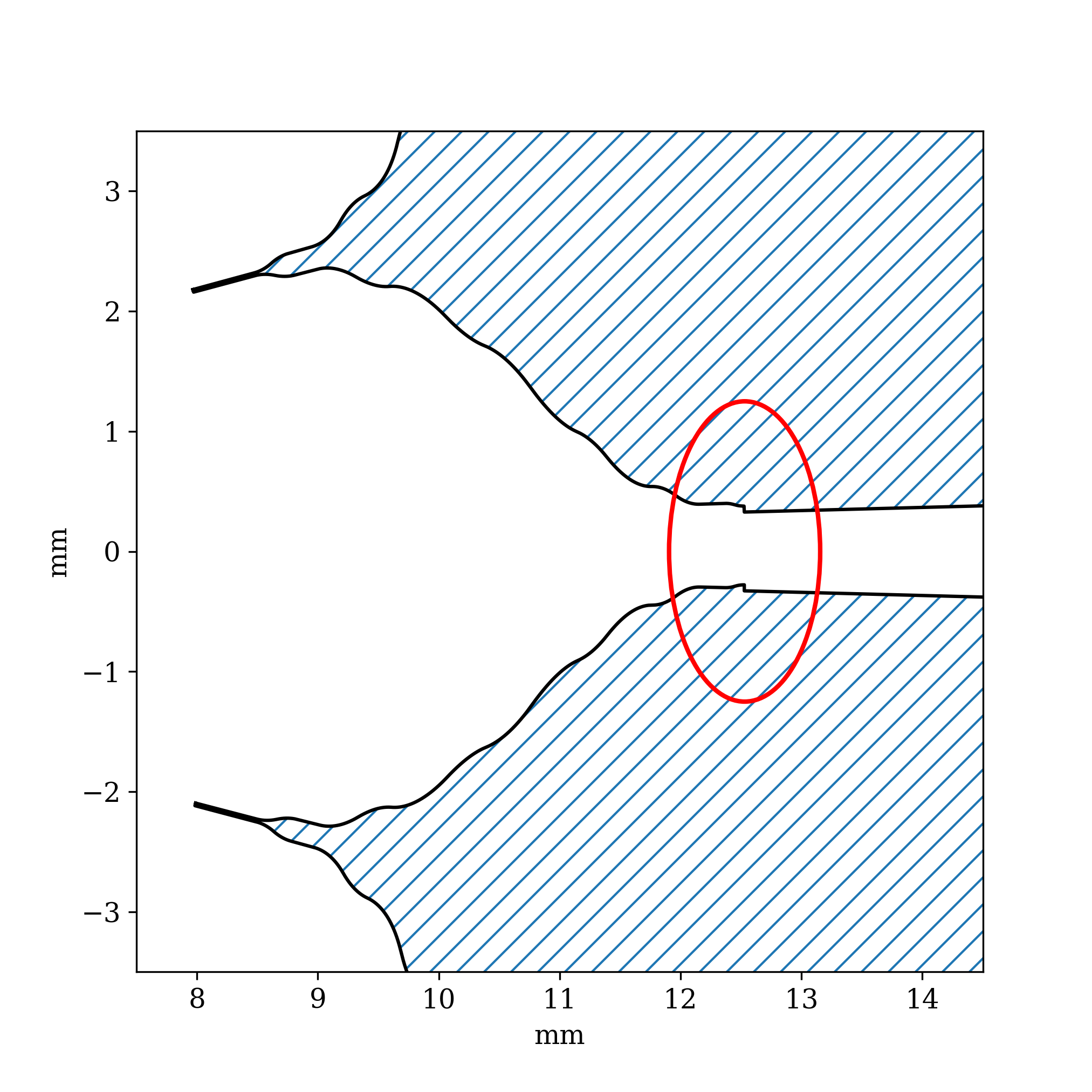}
\end{tabular}
\end{center}
\caption
{ \label{fig:tips}
A closeup of one petal in the design of mask M12P3. The inner starshade is rotated relative to the outer, creating exposed tips at all petals (seen at 12.5 mm). The blue hatched area is transparent; the white area is the silicon wafer.}
\end{figure}

In order to suspend the fragile silicon starshade in the testbed, the starshade design incorporates radial struts that keep the starshade attached to the outer supporting wafer. This means the end of the petal never terminates at a tip. The diffraction equations show that the critical features of the starshade are the inner valleys between the petals and the outer tips of the petals, i.e., where the petal shape has a large azimuthal component. To verify that critical features such as tips behave in an expected manner, and to make the test article reflect a more flight-like configuration, we tested a mask built with exposed tips. Figure~\ref{fig:tips} shows the design of mask M12P3 where the inner starshade is slightly rotated relative to the struts and outer apodization function to expose tips at the end of the petals. As the petalized starshade is an approximation to a radial, azimuthally-symmetric apodization profile, rotating the inner starshade does not change the approximated radial apodization profile and therefore should achieve the same contrast.

\begin{figure}[htb]
\begin{center}
\begin{tabular}{c}
  \includegraphics[width=0.9\linewidth]{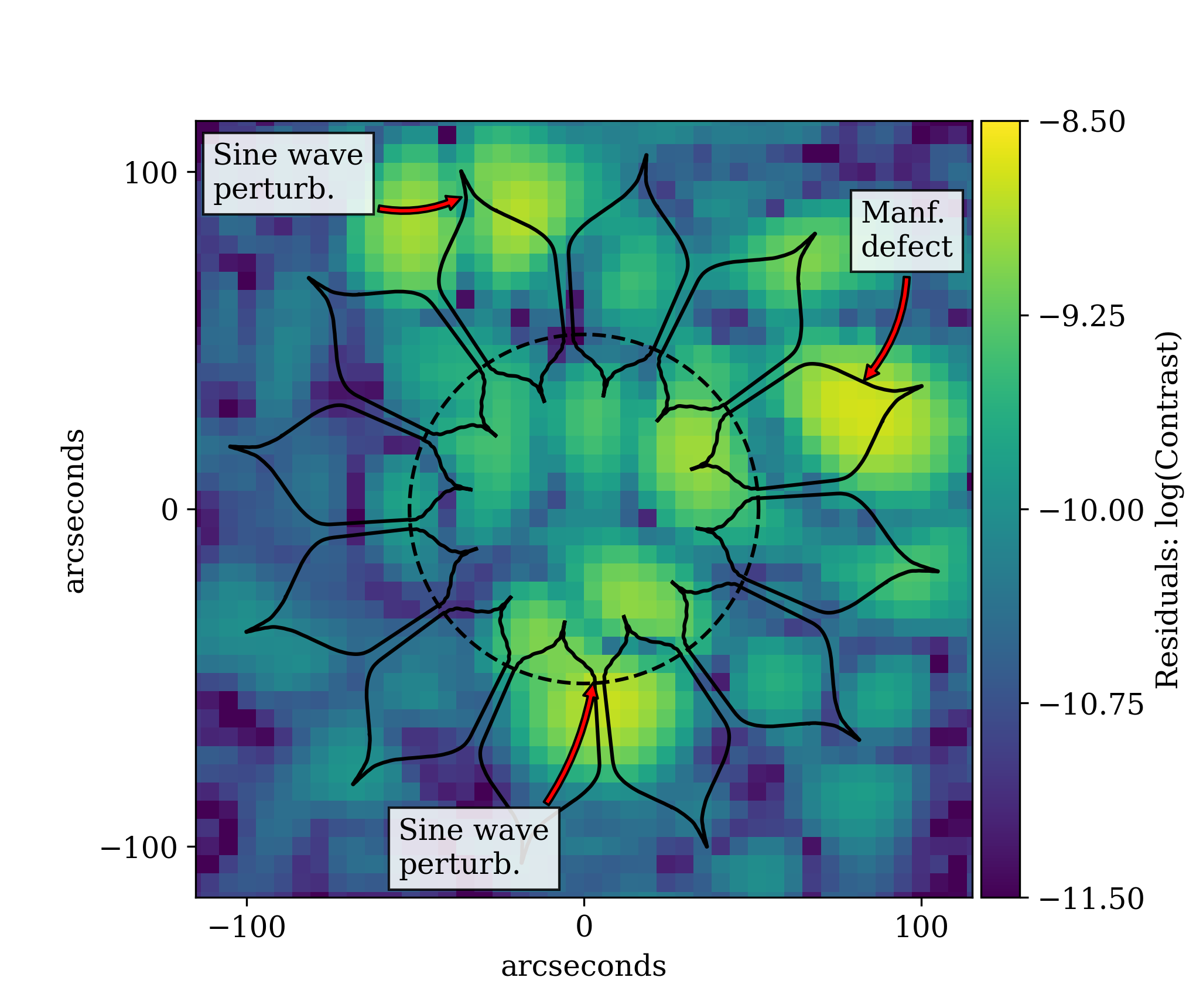}
\end{tabular}
\end{center}
\caption
{ \label{fig:tip_contrast}
Residuals between experiment and model of stacked [log-scale contrast] images of mask M12P3 at $\lambda~=~725$~nm. Lab and model images were created by median combining images taken at three starshade orientations. The brightest spots are due to intentional sine wave perturbations (see Sec.~\ref{sec:sine_waves}) and an unintentional manufacturing defect. High contrast is still achieved at the location of the exposed tips (marked by the dashed circle).}
\end{figure}

Figure~\ref{fig:tip_contrast} shows the residuals between experiment and model for a stacked image of mask M12P3 at $\lambda=725$~nm. We imaged the starshade rotated by 0$^\circ$, 120$^\circ$, and 240$^\circ$, de-rotated the images to align on the perturbations, and then median combined. Imaging at different orientations lessens the impact of the central polarization lobes. The residuals ($|$experiment $-$ model$|$) of the stacked images are shown in Fig.~\ref{fig:tip_contrast}. In addition to the exposed tips, M12P3 has sine wave perturbations built into its shape for model validation (see Sec.~\ref{sec:sine_waves}) and a defect leftover from the manufacturing process, which dominate the contrast in the residual image.

\new{Taking a $\lambda/D$ wide annulus at the angular separation of the tips, and excluding the portion that lies on the sine wave perturbation (at 6:00 in the image), we find the average residual contrast in the annulus is $7\times10^{-11}$.} Inspection of Fig.~\ref{fig:tip_contrast} shows there is not significant residual light at the location of the tips and that most of the residual is due to leakage of unmodeled non-scalar diffraction from the inner gaps of the starshade and from the perturbations. From this we conclude that the exposed tips behave as expected (to the 10$^{-10}$ contrast level) and the starshade still provides sufficient contrast outside the IWA.

\section{Model Validation Experiments}
\label{sec:validation_experiments}
There will always be slight deformations in the starshade's shape due to manufacturing errors, deployment errors, and thermal and mechanical stresses. A contrast error budget sets tolerances on the allowed deformations by balancing aspects of the mechanical design deemed to be most challenging\cite{Shaklan_2015}. \new{One purpose of the model validation experiments is to validate the accuracy at which the models used to derive the error budget capture the sensitivity of contrast performance to shape perturbation; by observing starshades with known perturbations built into the shape, we can validate how the contrast changes in a known way. The validation accuracy set by the experiments determines the Model Uncertainty Factor (MUF) between contrast and shape in the error budget. The MUF is a multiplicative term applied to the intensity (contrast) of each term in the error budget that provides a margin for inaccuracy of the model. Reducing the MUF through experimentation allows us to reduce the contrast margin budgeted to model uncertainty and leads to a more efficient design.} The shape changes selected for the validation experiments are related to the mechanical architecture of the SRM design\cite{WFIRST_SRM} and are representative of the shape errors in the error budget\cite{Shaklan_2015}.

The model we are trying to validate uses scalar diffraction only, which is believed to be sufficiently accurate for the flight-scale starshade. However, since the discovery of non-scalar, thick screen effects in the lab configuration, additional work must be done to include these effects in the model in order to properly demonstrate that the perturbations are accounted for at flight scales.

In this section, we first present results from two experiments testing perturbations that dominate the error budget; experiments on other perturbations are in progress\cite{S5_Plan} and will be reported at a later date. For the perturbed shape tests, the perturbations are made large enough so that the predicted light leakages are dominated by scalar diffraction, with only slight contributions from the thick screen effect in the inner parts of the starshade. Later in this section, we present results from a polarization study demonstrating the non-scalar model accurately captures the thick screen effect. Estimates of this model applied to flight-scale starshades shows that the effect is negligible as expected (see Appendix~\ref{apx:skin_scaling}), but a full analysis is saved for a later date. We end this section with an experiment testing the Fresnel number dependence on contrast by testing a starshade at a Fresnel number outside of its designed Fresnel space. We show the model accurately captures this transition, as well as the transition between the scalar and non-scalar diffraction dominated regimes. \new{The results presented in this section are progress towards Milestone 2 of the S5 Project, which will satisfy the final requirement in the Starlight Suppression technology development plan\cite{S5_Plan}.}

\subsection{Optical model summary}
Several methods are available for solving the scalar diffraction problem, all of which have been shown to be in agreement. Two methods, a boundary diffraction wave method\cite{Cady_2012} and a similar angular integral method\cite{Cash_2011, Harness_2018}, convert the two-dimensional diffraction equation into a one-dimensional line integral around the occulter's edge and are well suited to capture the large dynamic range of sizes in the starshade shape. In this work, we use the model of Ref.~\citenum{Harness_2020}, which uses a two-dimensional Fresnel propagator (shown to be in agreement with the boundary methods\cite{Harness_2018}) to include non-scalar diffraction. We provide a brief summary of the optical model in Appendix~\ref{apx:model_description} and we show in Sec.~\ref{sec:polarization} that our implementation of non-scalar diffraction agrees with the data.

\subsection{Perturbed shape experiments}
\label{sec:perturbation_experiments}
\begin{figure}[htb]
\begin{center}
\begin{tabular}{c}
  \includegraphics[width=0.8\linewidth]{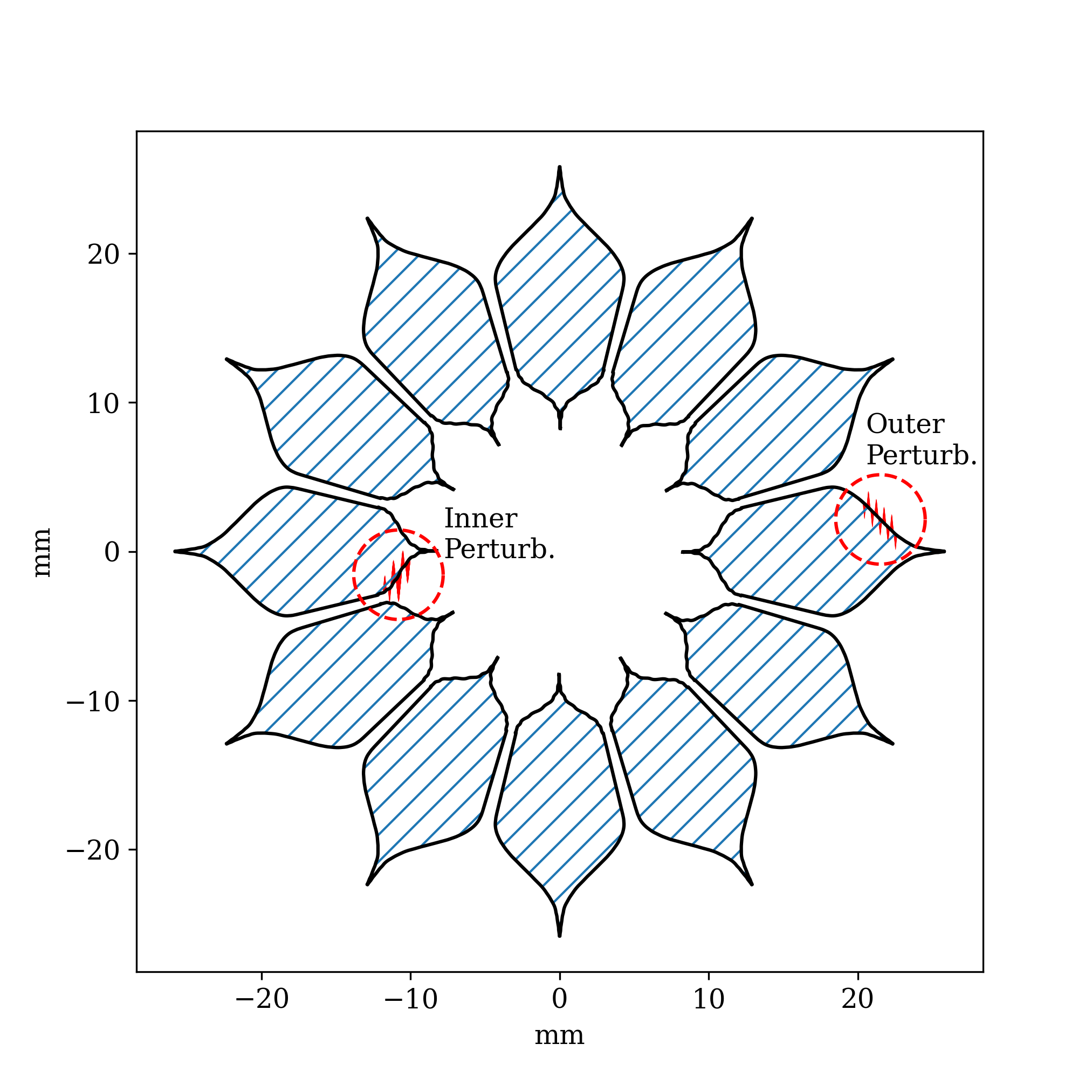}
\end{tabular}
\end{center}
\caption
{ \label{fig:defect_occulter}
Location of inner and outer perturbations on the perturbed masks M12P2 and M12P3. The red arrows are a quiver plot of the sine waves on M12P3 with amplitudes greatly exaggerated. The blue hatched area is transparent; the white area is the silicon wafer.}
\end{figure}

Here we present results from testing two masks with different classes of perturbations: displaced edges and sine waves. Each mask has two perturbations of different sizes built into its design. The sizes are chosen to produce a signal in the image that is bright enough to overcome contributions from the thick screen effect, but faint enough to be informative to model validation. One perturbation is located on the inner starshade and is made brighter since it is closer to the central polarization lobes; the other defect is located on the outer starshade and is allowed to be fainter. The perturbed masks have 12 petals to minimize the thick screen effect; fewer gaps between petals mean there are fewer sources of non-scalar diffraction and the lobes are $(12/16)^2$ times as bright. Details of the perturbations are presented in Table~\ref{tab:perturbations} and the locations of inner and outer perturbations are shown in Fig.~\ref{fig:defect_occulter}.

In comparing contrasts between experiment and model, we draw a photometric aperture of radius $\lambda/D$ around the perturbation and calculate the average contrast in that aperture using Eq.~(\ref{eq:average_contrast}) of Appendix~\ref{apx:contrast_definition}, with error given by Eq.~(\ref{eq:contrast_error}) of Appendix~\ref{apx:contrast_definition}. The expected contrast presented in Table~\ref{tab:perturbations} is the average contrast in the photometric aperture, calculated under the assumption of scalar diffraction only.
\begin{table}[ht]
\caption{Description of shape perturbations, including their expected contrast (photometric average, see Eq.~(\ref{eq:average_contrast})) at two wavelengths. Each perturbed mask ($^\mathrm{a}$M12P2, $^\mathrm{b}$M12P3) has an inner and outer perturbation of different size.}
\label{tab:perturbations}
\begin{center}
\begin{tabular}{ c c c c c }
\hline
\rule[-1ex]{0pt}{3.5ex}\multirow{2}{*}{Perturbation} & \multirow{2}{*}{Location}  & \multirow{2}{*}{Description} & \multicolumn{2}{c}{Expected Mean Contrast}\\
\rule[-1ex]{0pt}{3.5ex} & & & $\lambda=641$ nm & $\lambda=725$ nm\\
\hline\hline
\rule[-1ex]{0pt}{3.5ex}Displaced Edge & Inner Petal$^\mathrm{a}$ & 3.7\micron tall, 414\micron long & 2.3$\times10^{-9}$  & 1.7$\times10^{-9}$\\
\rule[-1ex]{0pt}{3.5ex}Displaced Edge & Outer Petal$^\mathrm{a}$ & 2.4\micron tall, 532\micron long & 6.5$\times10^{-10}$ & 7.4$\times10^{-10}$\\
\rule[-1ex]{0pt}{3.5ex}Sine Wave & Inner Petal$^\mathrm{b}$ & 1.75\micron amp., 4 cycles over 2.9 mm & 1.9$\times10^{-9}$ & 1.4$\times10^{-9}$\\
\rule[-1ex]{0pt}{3.5ex}Sine Wave & Outer Petal$^\mathrm{b}$ & 1.75\micron amp., 5 cycles over 2.3 mm & 9.6$\times10^{-9}$ & 7.3$\times10^{-10}$\\
\hline
\end{tabular}
\end{center}
\end{table}

\subsubsection{Displaced edge segments}
\begin{figure}[htb]
\begin{center}
\begin{tabular}{c}
  \includegraphics[width=0.8\linewidth]{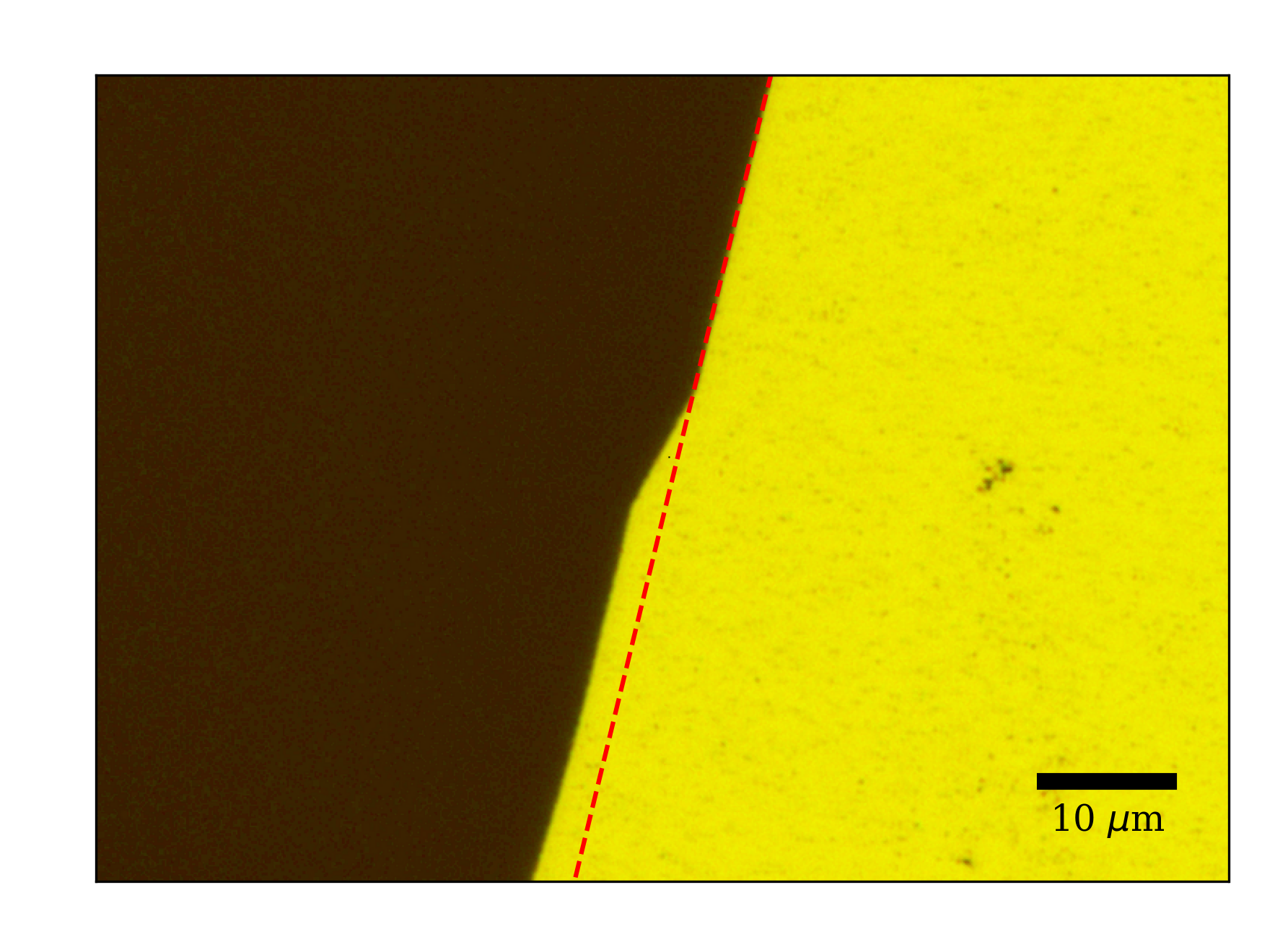}
\end{tabular}
\end{center}
\caption
{ \label{fig:displaced_image}
 Microscope image of the inner displaced edge perturbation on M12P2. The red line denotes the edge of the nominal starshade shape. The 3.7\micron step in the edge simulates a displaced edge segment.}
\end{figure}

The displaced edge perturbation simulates the effect of a petal edge segment being displaced from its nominal position during petal assembly on the ground. Figure~\ref{fig:displaced_image} shows a 3.7\micron tall displaced edge segment built into the design of the manufactured mask M12P2. The locations of the perturbations on the starshade petals are shown in Fig.~\ref{fig:defect_occulter}.

\begin{figure}
\begin{center}
\begin{tabular}{c}
  \includegraphics[width=.92\linewidth]{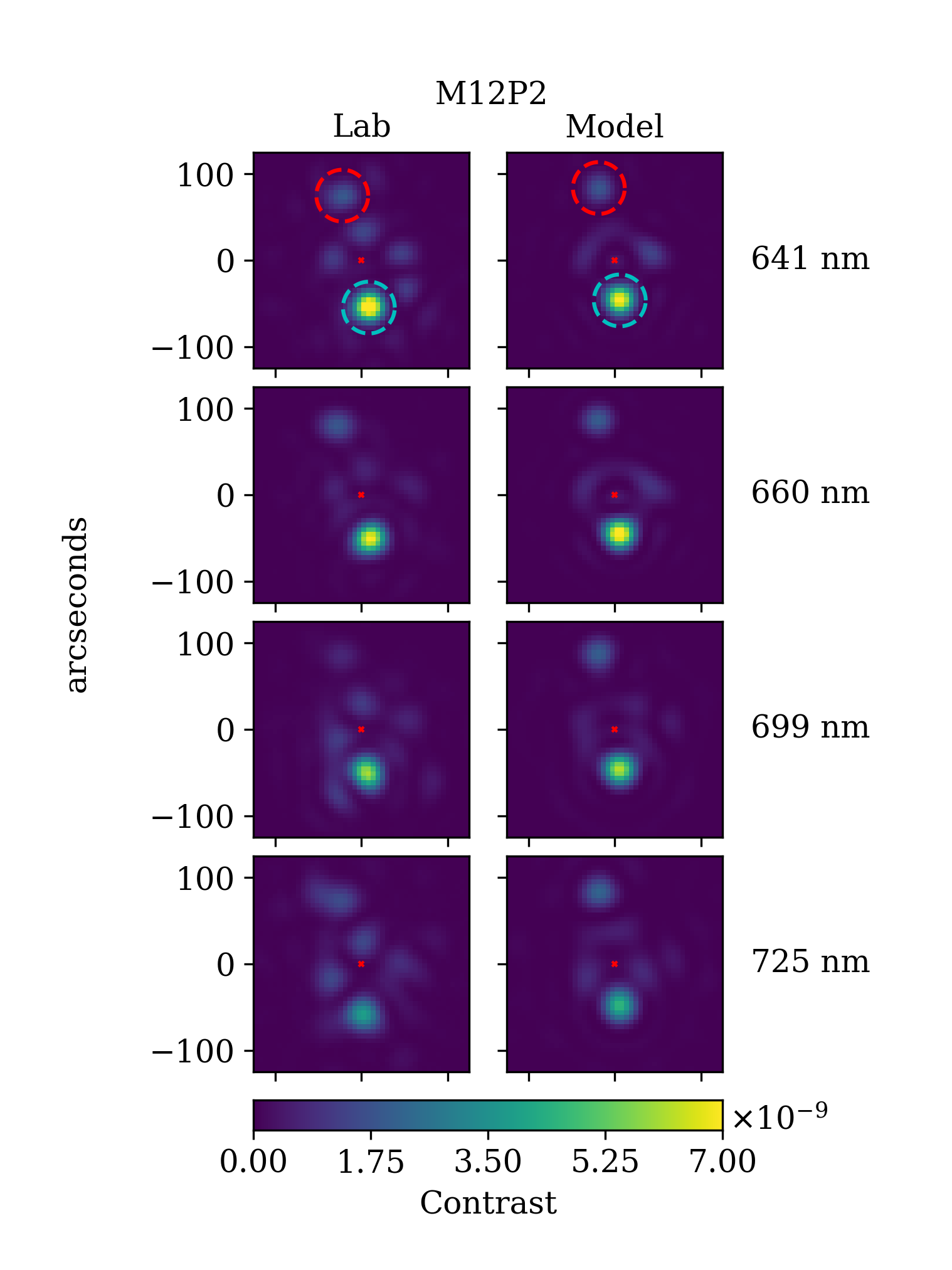}
\end{tabular}
\end{center}
\caption
{ \label{fig:m12p2_image}
 Contrast images of M12P2 $-$ displaced petal edge segment perturbations. Experimental data are in the left column, model data are in the right column. The rows are for the different wavelengths. The perturbation on the inner petal is circled in cyan and the perturbation on the outer petal is circled in red. The red `x' marks the center of the starshade.}
\end{figure}

\begin{figure}[htb]
\begin{center}
\begin{tabular}{c}
  \includegraphics[width=0.9\linewidth]{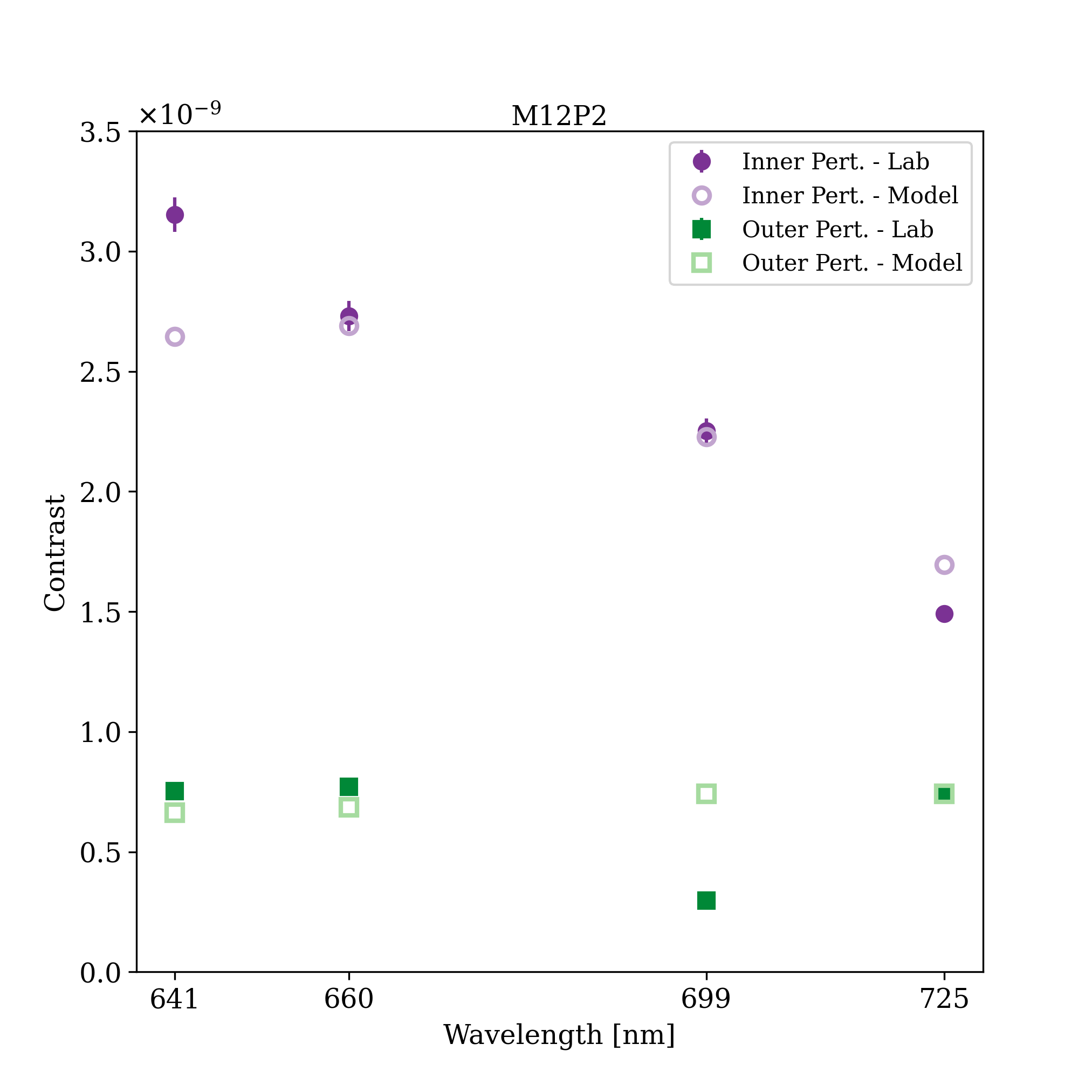}
\end{tabular}
\end{center}
\caption
{ \label{fig:m12p2_contrasts}
 M12P2 $-$ displaced edge perturbations. Contrast for the Inner and Outer perturbations across the four wavelengths. Solid markers are experimental data; open markers are model data. The contrast is the photometric aperture average given by Eq.~(\ref{eq:average_contrast}) and the error bars (on experimental data only) are $1\sigma$, given by Eq.~(\ref{eq:contrast_error}). \new{Error bars on experimental data that are not visible are smaller than the symbols.}}
\end{figure}

Figure~\ref{fig:m12p2_image} shows the experimental and model contrast images at four wavelengths across the bandpass. The model includes polarization effects and manufacturing defects found on the mask that are larger than 30 square microns. The input polarization vector is horizontal in the image so that the perturbations are away from the inner polarization lobes. Figure~\ref{fig:m12p2_contrasts} plots the average contrast in a $\lambda/D$ radius photometric aperture centered on each perturbation. By design, the inner perturbation is brightest and drops in contrast with wavelength. The model shows good agreement with the experimental data.

\begin{figure}[htb]
\begin{center}
\begin{tabular}{c}
  \includegraphics[width=0.9\linewidth]{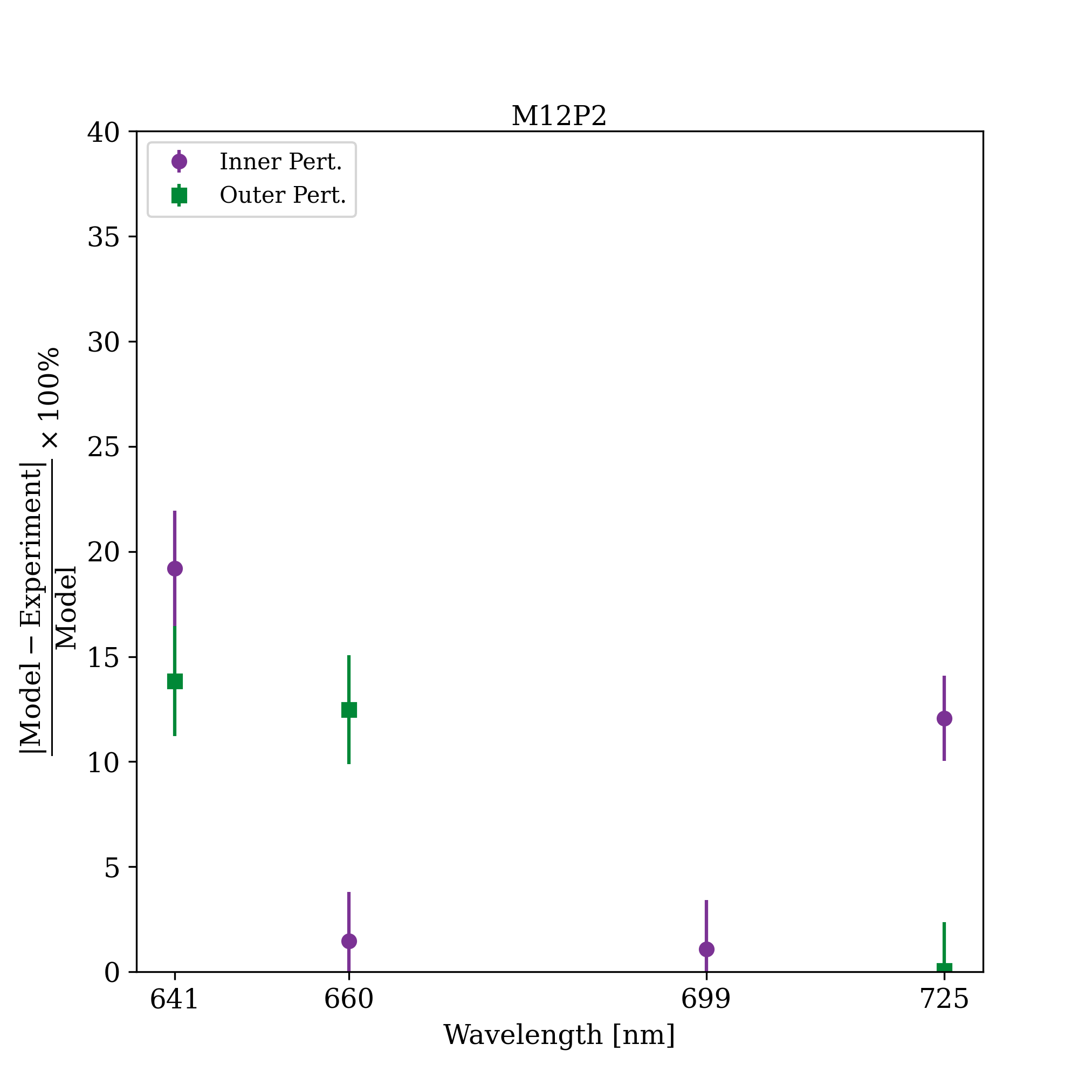}
\end{tabular}
\end{center}
\caption
{ \label{fig:m12p2_comparison}
 M12P2 $-$ displaced edge perturbations. Percent difference between experimental and model contrast for the Inner and Outer perturbations across the four wavelengths. The contrast is the photometric aperture average given by Eq.~(\ref{eq:average_contrast}) and the error bars are the percent uncertainty given by Eq.~(\ref{eq:percent_uncertainty}). The value for the Outer perturbation at 699~nm is off the chart (at 60\%).}
\end{figure}

Figure~\ref{fig:m12p2_comparison} plots the difference in the photometric aperture averaged contrast between the experimental and model data for each of the perturbations. The percent difference is calculated as
\begin{equation}
	\mathrm{Percent}~\mathrm{Difference} = \frac{\left|\mathrm{Model} - \mathrm{Experiment}\right|}{\mathrm{Model}} \times 100\%\,,
	\label{eq:percent_difference}
\end{equation}
where the values for Model and Experiment are the photometric aperture averaged contrast. The error bars in Figure~\ref{fig:m12p2_comparison} are the experimental uncertainty propagated to the percent difference, and are calculated as
\begin{equation}
    \mathrm{Percent}~\mathrm{Uncertainty} = \frac{\sigma_\mathrm{Experiment}}{\mathrm{Model}} \times 100\%\,,
    \label{eq:percent_uncertainty}
\end{equation}
where $\sigma_\mathrm{Experiment}$ is calculated from Eq.~(\ref{eq:constant_error}) of Appendix~\ref{apx:contrast_definition}.

The agreement is better than 20\% for all perturbations except for the outer perturbation at $\lambda=699$~nm. The difference for the outer perturbation at 699 nm is $\sim60\%$; Fig.~\ref{fig:m12p2_image} shows that it is much dimmer than expected and does not follow the trend of the different wavelengths. This effect is seen in three orientations of the mask (rotated by $\pm$ 120$^\circ$) and has been repeated several times. We currently do not have a good explanation as to why this happens, but we are continuing to refine the characterization and model of the optical edges in hopes of explaining this observation.

\subsubsection{Sine waves}
\label{sec:sine_waves}
Sinusoidal changes to the edge shape can occur if individual edge segments are misplaced in such a way that their envelope creates a sine wave with respect to the nominal edge position. This is a particularly harmful perturbation if the sine wave is in sync with the Fresnel half-zones and they constructively interfere. This also places a strong wavelength dependence on the contrast they induce. The locations of the sine wave defects are shown in Fig.~\ref{fig:defect_occulter}.
\begin{figure}
\begin{center}
\begin{tabular}{c}
  \includegraphics[width=0.92\linewidth]{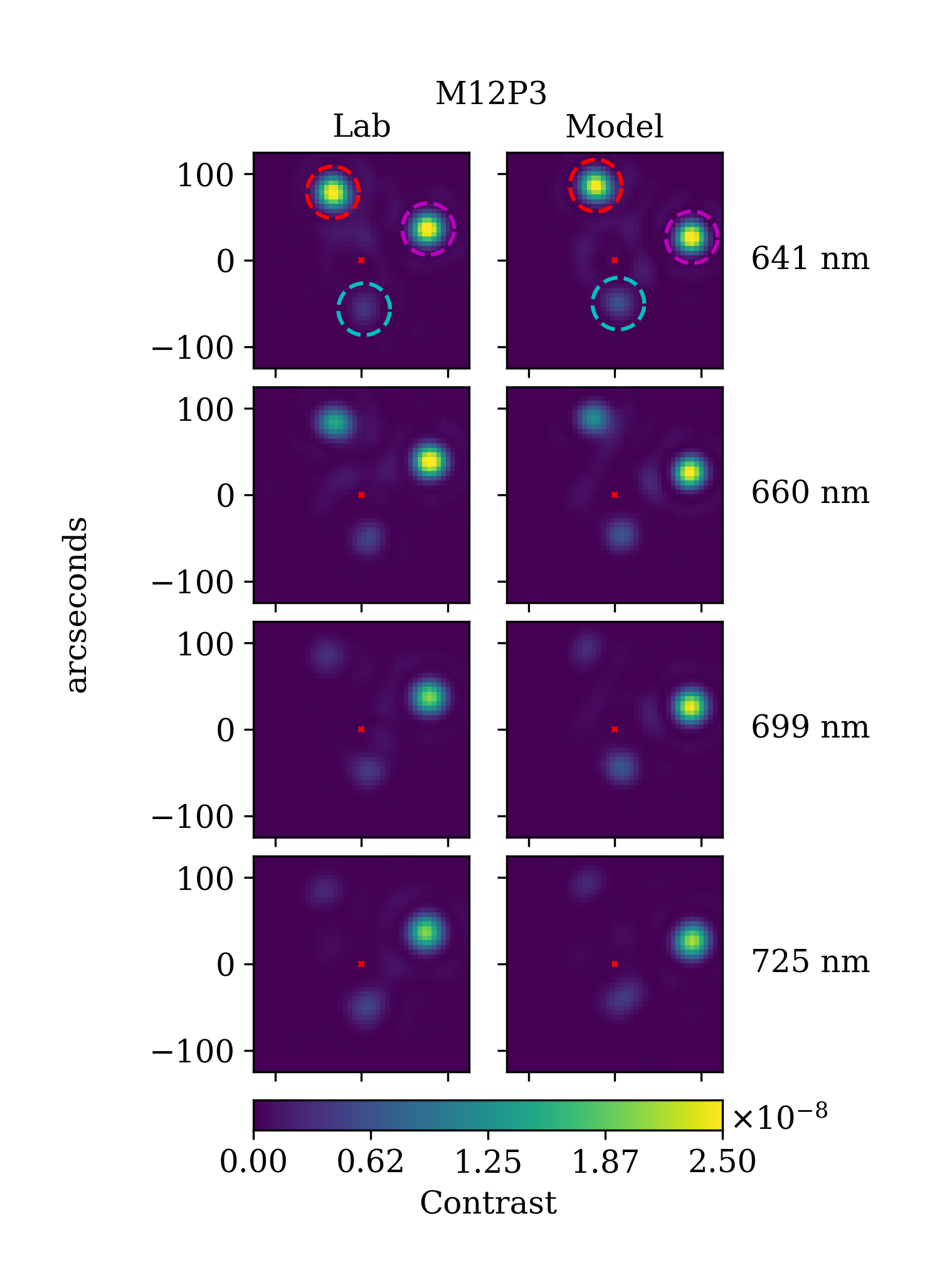}
\end{tabular}
\end{center}
\caption
{ \label{fig:m12p3_image}
 Contrast images of M12P3 $-$ sine wave perturbations. Experimental data are in the left column, model data are in the right column. The rows are for the different wavelengths. The perturbation on the inner petal is circled in cyan, the perturbation on the outer petal is circled in red, and the unintentional manufacturing defect is circled in magenta. The red `x' marks the center of the starshade.}
\end{figure}

Figure~\ref{fig:m12p3_image} shows the images taken of this perturbed mask. For M12P3, in addition to the intentional perturbations, a large defect is left over from the manufacturing process and is the brightest source in the image. SEM images show this is a large defect that extends vertically below the wafer's device layer and has a complicated, protruding structure. We estimate the area to be $\sim$~750~$-$~1500 square microns, but projection effects make it difficult to know how much is seen by the camera. In the model, we adjust the area of the defect to 1375 square microns in order to match the data and contribute the appropriate amount of leakage to the other perturbations.
\begin{figure}[htb]
\begin{center}
\begin{tabular}{c}
  \includegraphics[width=0.9\linewidth]{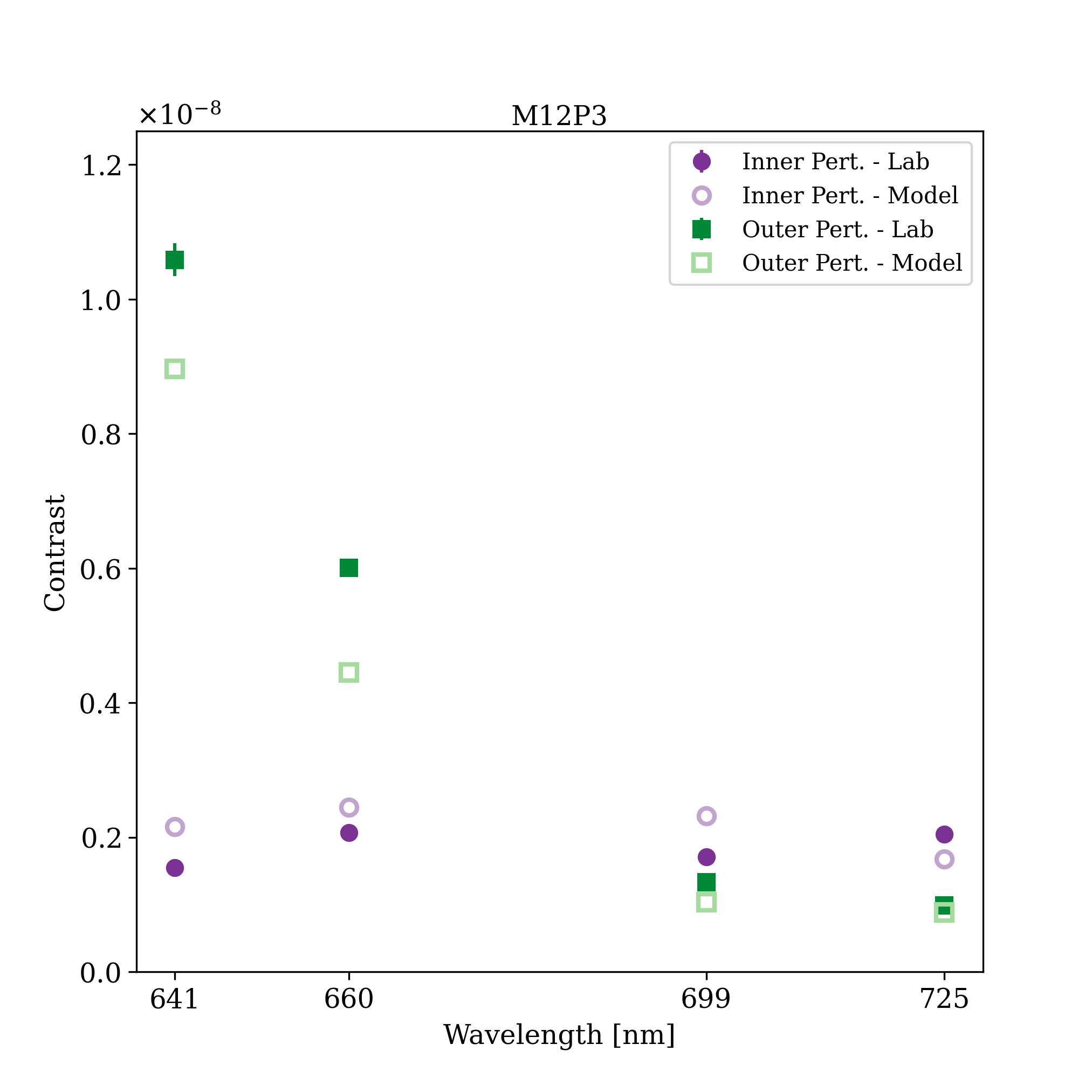}
\end{tabular}
\end{center}
\caption
{ \label{fig:m12p3_contrasts}
 M12P3 $-$ sine wave perturbations. Contrast for the Inner and Outer perturbations across the four wavelengths. Solid markers are experimental data; open markers are model data. The contrast is the photometric aperture average given by Eq.~(\ref{eq:average_contrast}) and the error bars (on experimental data only) are $1\sigma$, given by Eq.~(\ref{eq:contrast_error}). \new{Error bars on experimental data that are not visible are smaller than the symbols.}}
\end{figure}

\begin{figure}[htb]
\begin{center}
\begin{tabular}{c}
  \includegraphics[width=0.9\linewidth]{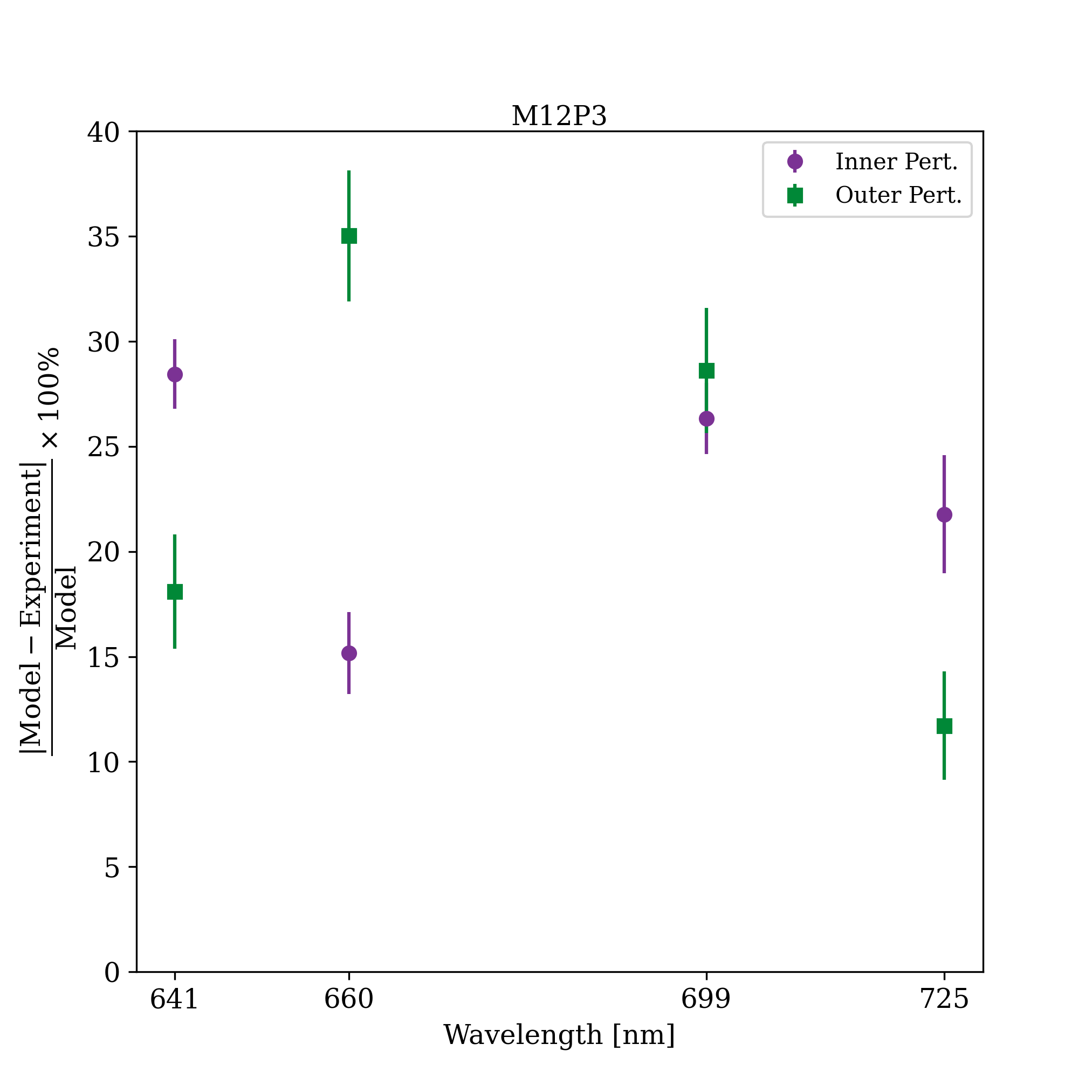}
\end{tabular}
\end{center}
\caption
{ \label{fig:m12p3_comparison}
 M12P3 $-$ sine wave perturbations. Percent difference between experimental and model contrast for the Inner and Outer perturbations and the unintentional manufacturing defect across the four wavelengths. The contrast is the photometric aperture average given by Eq.~(\ref{eq:average_contrast}) and the error bars are the percent uncertainty given by Eq.~(\ref{eq:percent_uncertainty}).}
\end{figure}

Figure~\ref{fig:m12p3_contrasts} plots the average contrast in a photometric aperture centered on each perturbation. The sine wave perturbation has a strong response with wavelength, and the inner and outer perturbations occasionally switch which is brightest, behavior that agrees with model predictions. Figure~\ref{fig:m12p3_comparison} shows the comparison between the experimental and model contrasts, with the percent difference calculated from Eq.~(\ref{eq:percent_difference}) and the uncertainty on that difference calculated from Eq.~(\ref{eq:percent_uncertainty}). For this mask, since the inner and outer perturbations are of equal brightness, the inner one generally performs worse as it is closer to and suffers more contamination from the central polarization lobes. Both the inner and outer perturbations agree with the model to better than 35\% at all wavelengths. This agreement is not as good as that of the displaced edges mask, which we attribute to the presence of the large manufacturing defect. Due to the complicated, three-dimensional structure of the manufacturing defect, it is difficult to model the interaction between the defect and the sine wave perturbations. We believe this unmodeled interaction is the source of the larger discrepancy.

\subsection{Polarization study}
\label{sec:polarization}
\begin{figure}[htb]
\begin{center}
\begin{tabular}{c}
  \includegraphics[width=\linewidth]{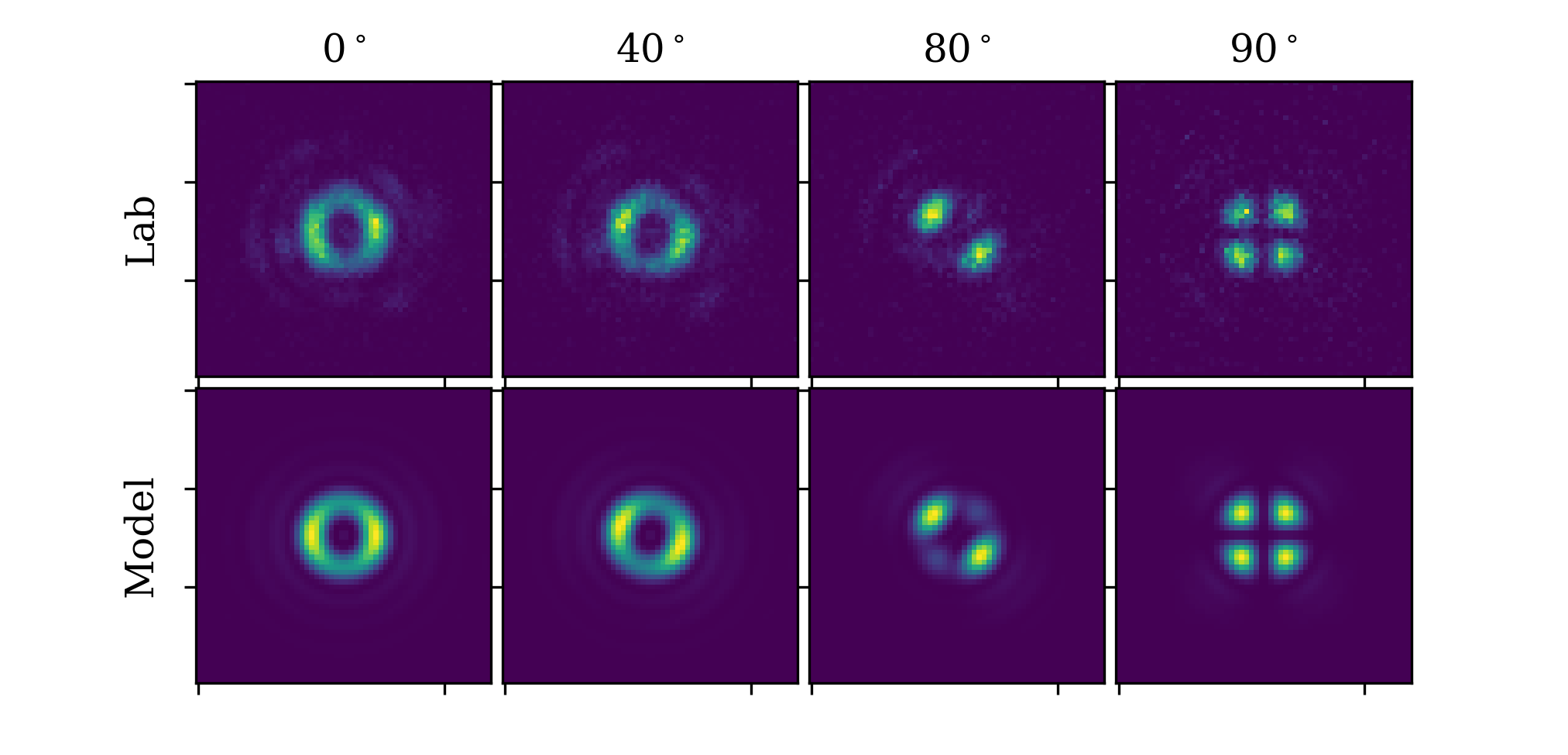}
\end{tabular}
\end{center}
\caption
{ \label{fig:rot_img}
  Experimental (top) and model (bottom) images of mask DW9 imaged with the camera analyzer rotating relative to the input polarizer direction (horizontal in the image). The angle given is that between the analyzer and polarizer, where 0$^\circ$ means they are aligned, 90$^\circ$ means they are crossed. Data are taken with $\lambda=641$~nm.}
\end{figure}

Our proposition that the bright central lobes are due to the thick screen effect can be validated by examining the polarization induced as light propagates past the mask. In the explanation provided in Sec.~\ref{sec:thick_screen}, the change in the electric field is dependent on the polarization direction relative to the sidewall of the mask. The effect is dominant in the inner gaps between petals, which can be approximated by parallel plates aligned with the clocking angle of the petal. Assuming horizontal linearly polarized light, petals at the 3:00 and 9:00 positions have all $s$-polarization (electric field parallel to wall), petals at 12:00 and 6:00 have all $p$-polarization (electric field perpendicular to wall), and petals in between have a mix of both. The $s$-polarization is subject to a greater change in the electric field (the field goes to zero at the walls of a perfect conductor\cite{Jackson}), so the polarization lobes are aligned with the input polarization direction. Figure~\ref{fig:rot_img} confirms this as we view horizontally polarized input light with an analyzing polarizer in front of the camera rotated to different angles. When the analyzer is exactly orthogonal to the input polarizer, scalar diffraction theory predicts no light should be visible. However, we see four lobes that are the result of unbalanced vertical polarization induced by the petals at $45^\circ, 135^\circ, 225^\circ, 315^\circ$.

\begin{figure}[htb]
\begin{center}
\begin{tabular}{c}
  \includegraphics[width=0.9\linewidth]{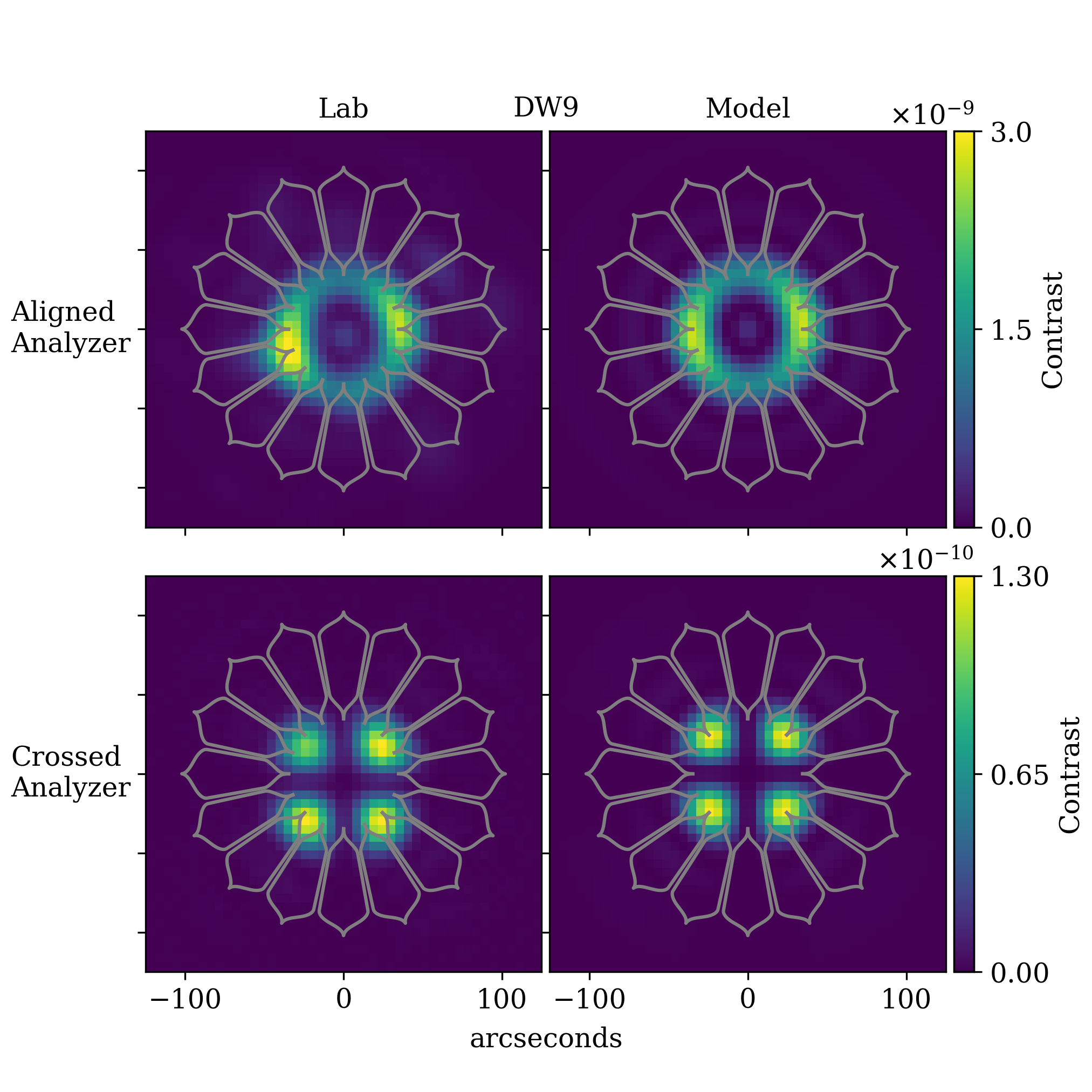}
\end{tabular}
\end{center}
\caption
{ \label{fig:dw9_comp}
 Experimental (left) and model (right) images of mask DW9 at $\lambda=641$~nm. In the top row, the camera analyzer is aligned with the input polarizer (with a slight 15$^\circ$ misalignment). In the bottom row, the camera analyzer is crossed with the input polarizer. {\bf Note} the difference in colorbar scales.}
\end{figure}

\begin{figure}[htb]
\begin{center}
\begin{tabular}{c}
  \includegraphics[width=0.9\linewidth]{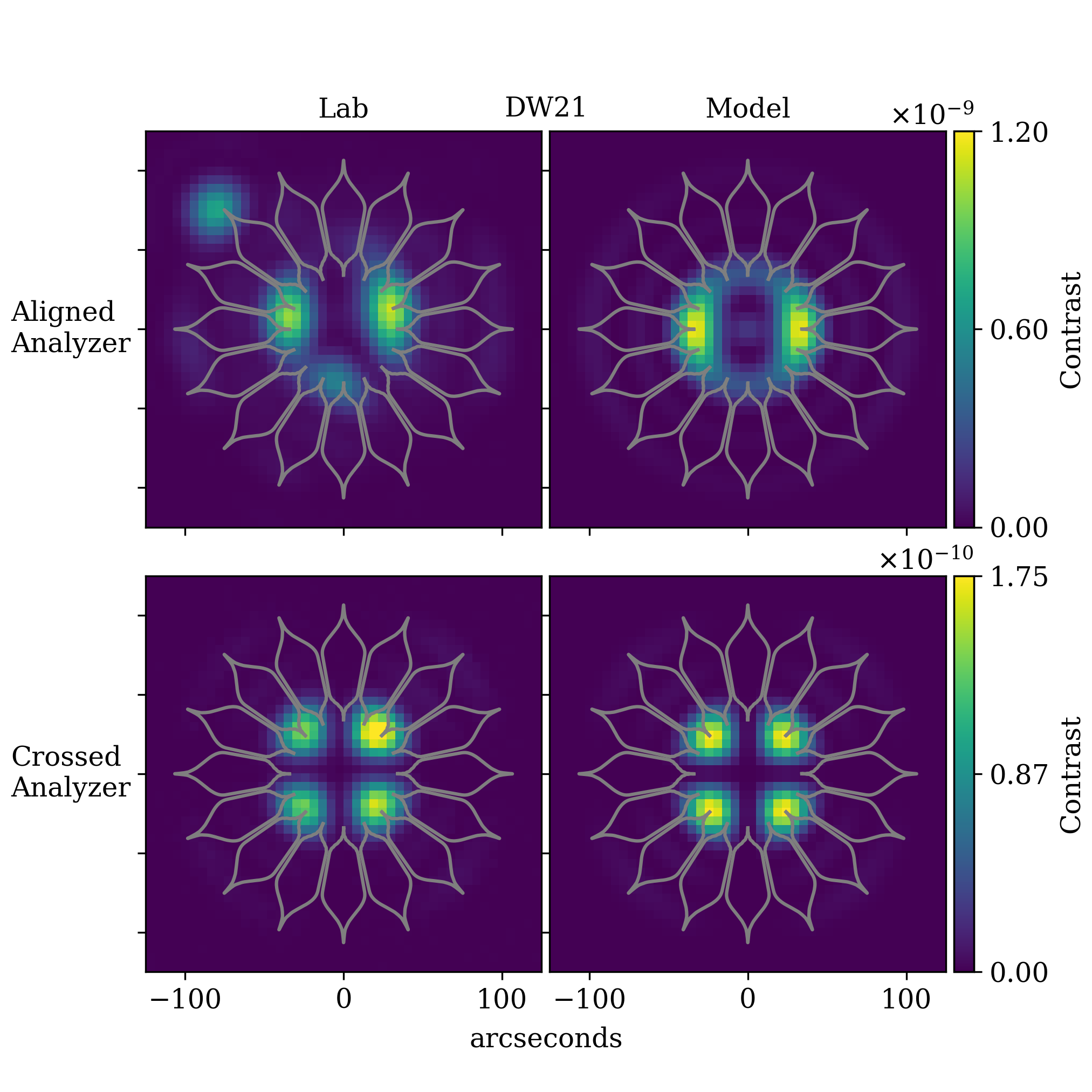}
\end{tabular}
\end{center}
\caption
{ \label{fig:dw21_comp}
 Experimental (left) and model (right) images of mask DW21 at $\lambda=641$~nm with polarized light. In the top row, the camera analyzer is aligned with the input polarizer. The bright spot at 10:00 is dust on the mask. In the bottom row, the camera analyzer is crossed with the input polarizer. {\bf Note} the difference in colorbar scales.}
\end{figure}

Figure~\ref{fig:dw9_comp} shows experimental and model data at $\lambda=641$~nm for the 7\micron thick mask DW9 with the analyzer aligned and crossed with the input polarization. The model agrees well with the experiment and is able to replicate the major polarization features, even at contrast levels below $10^{-10}$. \new{Figure~\ref{fig:dw21_comp} shows experimental and model data at $\lambda=641$~nm for the 3\micron thick mask DW21 (these data are taken with linear polarizer installed and thus differ from the DW21 data taken in Sec.~\ref{sec:broadband}). Again, the model agrees well and the peak contrast is $\sim$ 2.3$\times$ fainter than those of DW9, which is consistent with the thick screen effect scaling linearly with edge thickness, as should be expected. This demonstration shows we understand the source of the non-scalar diffraction, are able to replicate it's dependency on edge thickness, and that it is due solely to the small scale of the experiment and will not be an issue for flight. Appendix~\ref{apx:skin_scaling} provides an estimate of the strength of this effect at flight scales.}

\subsection{Contrast dependence on Fresnel number}
\label{sec:varying_fresnel}
\new{The sub-scale experiments presented here are informative for a starshade mission because of the scale invariance of the diffraction equations under the Fresnel approximation. Therefore it is critical to validate this assumption via an exploration of the Fresnel parameter space, starting in regions of high contrast and watching the contrast degrade as the experiment moves beyond the design space. Future testing should deconstruct the Fresnel number into its three constituent parts $\left(R,\lambda,Z\right)$ and explore the contrast dependence of each around the flight-like Fresnel number\cite{SSWG}, however, the size of this testbed puts a practical constraint on how much of the Fresnel parameter space can be explored. Section~\ref{sec:broadband} explored the wavelength dependence within the designed Fresnel space; in this section, we explore the size dependence and push beyond the designed Fresnel space to where the contrast begins to degrade. This experiment is a first step in an exploration of Fresnel space and should be followed by more extensive tests in the future.}

\begin{figure}[htb]
\begin{center}
\begin{tabular}{c}
  \includegraphics[width=\linewidth]{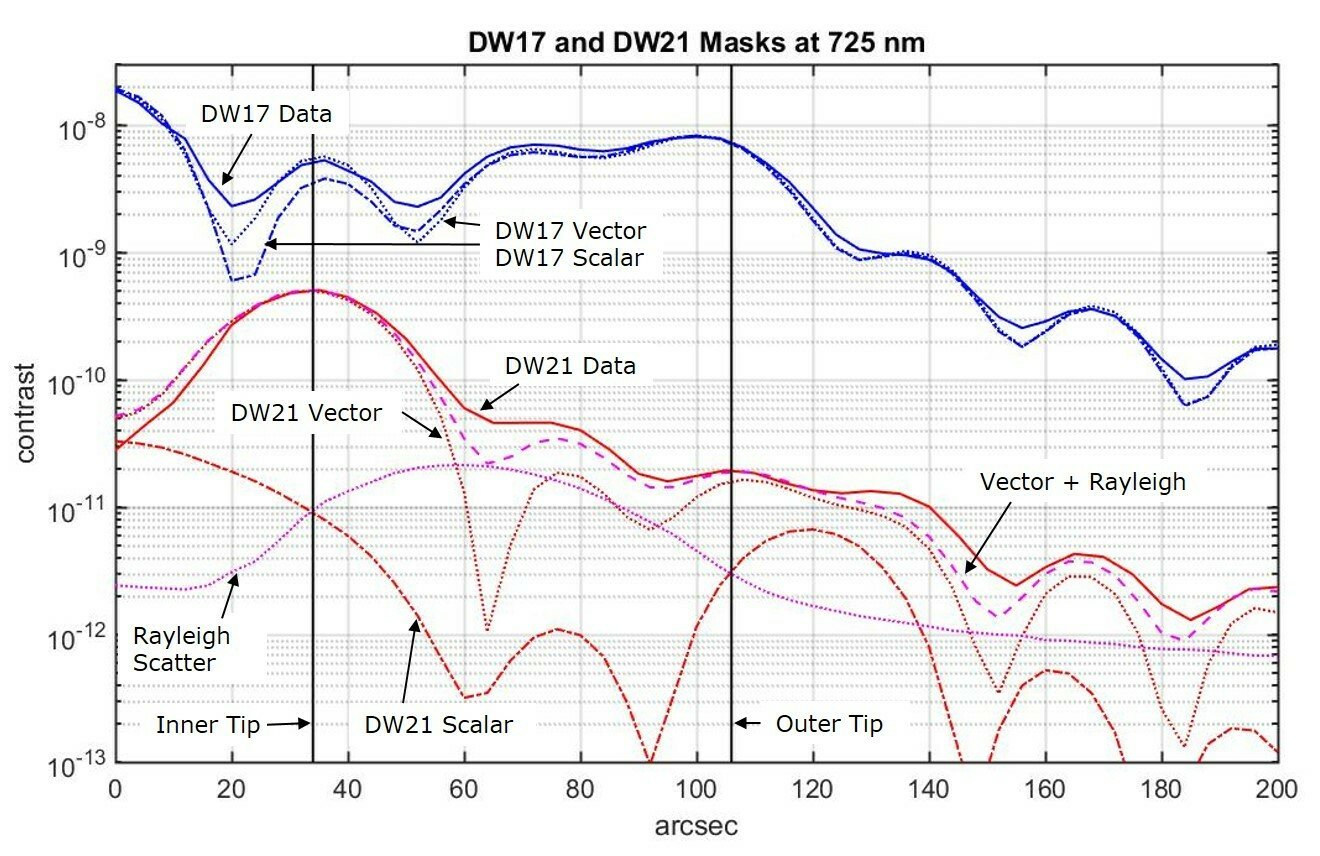}
\end{tabular}
\end{center}
\caption
{ \label{fig:out_of_band}
  Azimuthally averaged contrast for masks DW17 and DW21 at $\lambda=725$~nm, and the corresponding vector and scalar propagation models. The vertical lines mark the tip of the inner and outer apodization profiles. Also shown are estimates of Rayleigh scattering generated from the models of Ref.~\citenum{Willems_2019}.}
\end{figure}

Figure~\ref{fig:out_of_band} shows the results of testing a starshade outside of its designed Fresnel space, where the change in Fresnel number is achieved by changing the size of the starshade. Masks DW17 and DW21 have identical apodization functions, but DW21 is 3\% larger, which means $\lambda=725$~nm light is in the designed Fresnel space of DW21 $(N=12.2)$, but outside that of DW17 $(N=11.5)$. Figure~\ref{fig:out_of_band} shows the results of both masks at 725 nm; the contrast rises steeply with Fresnel number and the model captures this transition well, with even better model-experiment agreement at the brighter contrast. At these brighter contrast levels, the diffraction is completely dominated by scalar diffraction and both the scalar and non-scalar models converge. As the model agreement improves as the experiment moves beyond the regime where non-scalar diffraction is present and into a scalar-only regime, we build confidence that the scalar model will accurately predict performance in configurations where non-scalar diffraction is less prominent.

\section{Discussion}
\label{sec:discussion}

The starshade enables the detection of exoplanets because it provides high contrast at a small IWA, which means it operates at a moderately low Fresnel number where the diffraction is governed by complicated, near field equations. The apodization function that provides sufficiently high contrast is found by solving an optimization problem that minimizes the electric field in the Fresnel-Kirchhoff diffraction equation. This radial apodization function is then approximated by a petalized binary occulter that is the starshade's edge. Until now, it had not been shown that we have the tools capable of designing an apodization function that sufficiently suppresses diffraction at low Fresnel numbers and that the petalized approximation is valid. As the Fresnel number is decreased, diffraction becomes harder to control as the apodization function spans fewer Fresnel zones over which to average out imperfections. As such, it was necessary to demonstrate that high contrast was achievable at low Fresnel numbers. The experiments presented here show for the first time that the design tools used to solve for the apodization function are capable of achieving $10^{-10}$ contrast at a flight-like Fresnel number. The experiment contrast floor was $\sim10^{-10}$ at the IWA and $\sim10^{-11}$ near the outer starshade, allowing us to conclude that approximations in the model are no worse than $10^{-10}$ for the nominal starshade design.

The observed thick screen effect represents a worse case scenario in which our assumption of a purely scalar theory of diffraction is no longer valid and the optimized apodization is no longer applied to the appropriate problem. However, even at the small scales of the laboratory configuration, the contribution from non-scalar diffraction is below the target contrast level. Additionally, we expect deviations from scalar diffraction theory to go away at larger scales as the sizes of features get much larger than the wavelength; scaling to a larger starshade for flight should work in our favor. \new{If our theory of the thick screen is correct, the edges of the full-sized starshade will induce the same wavelength wide boundary layer near the edge, but the size of the starshade will be 1000$\times$ larger, meaning the impact of the non-scalar diffraction will be 10$^6\times$ smaller in intensity and will be reduced to a negligible amount (see Appendix~\ref{apx:skin_scaling}). Additional work may be needed to complete the validation of the non-scalar diffraction model, but the experiments completed thus far have built confidence that we can achieve the same, if not better, contrast at larger scales.} Given historic understanding of how light behaves around features with sizes comparable to the wavelength, it was known non-scalar diffraction could be present, but we previously did not have adequate tools to quantify the effect. The experiments in this work have helped to develop those tools and will allow us to apply them to future configurations.

The model validation experiments aim to determine the accuracy to which the model can predict the contrast sensitivity to known effects. Some of the experimental noise is due to unknown manufacturing errors, misalignment, turbulence, and stray reflections, and is not attributable to the model. Without detailed measurements of these error sources, we can only set an upper limit to the model inaccuracy. The results presented in Sec.~\ref{sec:perturbation_experiments} show the model remains accurate to at least the 35\% level, with an average difference of 20\%. The model agreement is even better ($<20\%$) for the displaced edge perturbation data on mask M12P2, as that mask did not have a large manufacturing defect as mask M12P3 did. This leads us to believe that the dominant source of model disagreement is due to uncharacterized perturbations in the manufactured mask, where due to the thick screen effect, the three-dimensional structure of the mask becomes relevant. The large manufacturing defect in M12P3 has a complex vertical structure that is difficult to adequately characterize and thus difficult to include in the model. Global defects, such as a variable overetching, are also difficult to characterize and can contribute to the model disagreement.

\begin{figure}[htb]
\begin{center}
\begin{tabular}{c}
  \includegraphics[width=\linewidth]{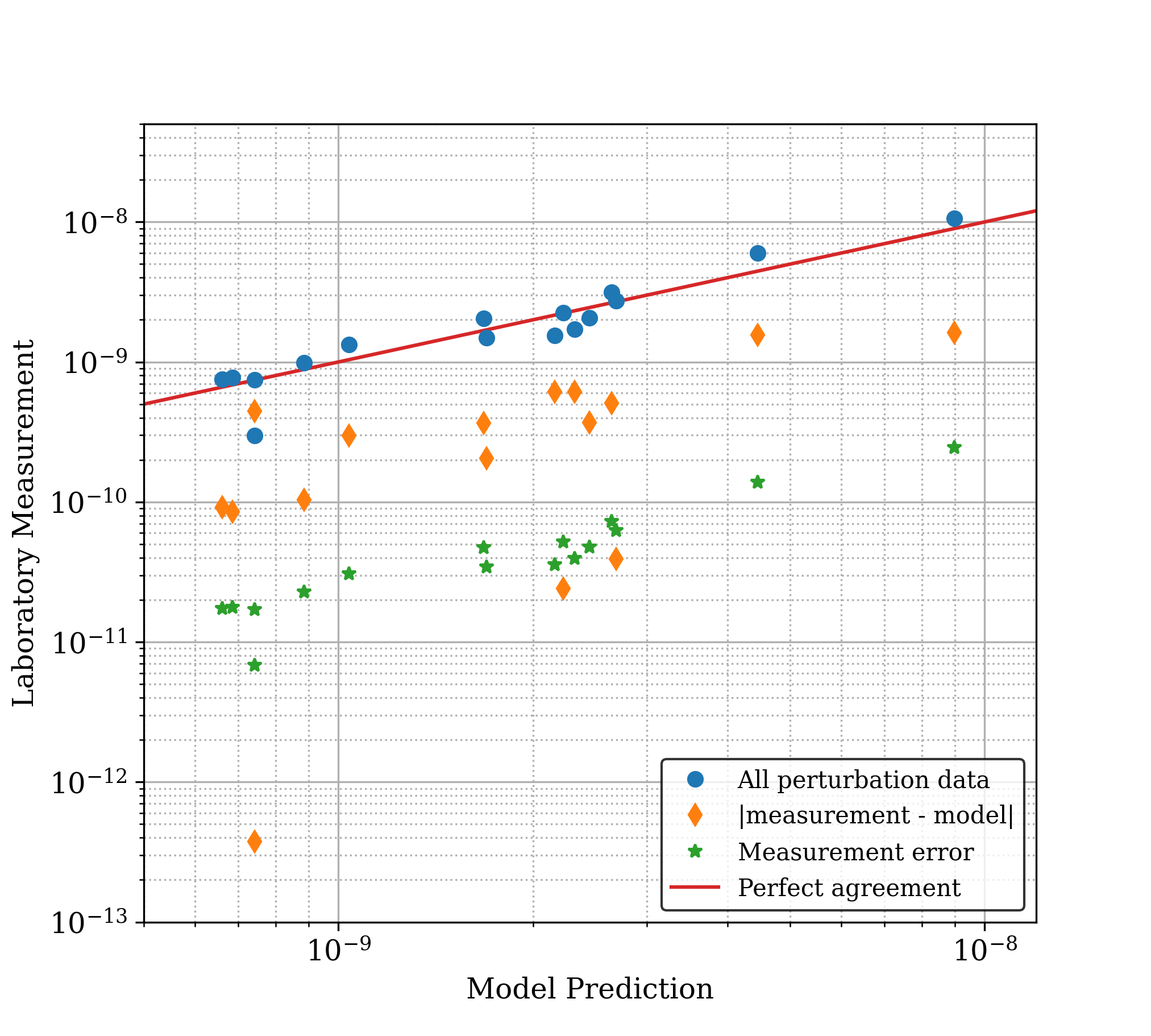}
\end{tabular}
\end{center}
\caption
{ \label{fig:model_v_lab}
Model vs. measured data for all perturbations on the displaced edge and sine wave perturbations and four wavelengths. The absolute difference between model and measured data and the measurement error are also shown.
}
\end{figure}

Figure~\ref{fig:model_v_lab} plots the measured contrast against the model prediction for all perturbation data, along with the absolute difference between the two and the measurement error. The measurements track well with the model, even down to the lowest contrast levels. The measurement-model difference and the measurement error follow the same trend as the measurements, meaning the fractional error is not increasing. \new{These results show that the model inaccuracy scales with the size of the perturbation and thus justifies the use of a multiplicative MUF. To provide sufficient margin in the error budget due to model uncertainty, the contrast from each perturbation must by increased by a factor of 1.35$\times$. While this conclusion certainly holds for perturbations down to 10$^{-9}$ contrast, due to the non-scalar diffraction present in the laboratory configuration, comment on the behavior of perturbations at lower contrast levels requires a cautious extrapolation of the trend seen in Fig.~\ref{fig:model_v_lab}.}

The experiments presented here are only the first of a suite of model validation experiments, so these conclusions on model accuracy are not complete. Future experiments\cite{S5_Plan} will include global and mixed perturbations and will help us further investigate coherent effects.

\section{Conclusion}
\label{sec:conclusion}
We have presented results from a number of experiments demonstrating the best contrast achieved with a starshade at a flight-like Fresnel number. We achieved better than $10^{-10}$ contrast, the level needed to detect Earth-like exoplanets, at the geometric IWA and across a scientifically interesting bandpass. The starshades tested were not perfectly manufactured, their small size introduced non-scalar diffraction, and the experiments were conducted in air. That we were able to achieve such high contrast even in the presence of these factors shows the efficiency and robustness in which starshades operate and should garner confidence in the community for a successful starshade mission in the future.

In future work, we will continue model validation experiments on test masks with more perturbations representative of those in flight. These include: displacing a single petal radially; displacing all petals radially; and combining an edge segment displacement with a petal displacement. We will also continue to refine the thick screen model through tests of masks with different thicknesses and formally apply that model to the flight design to show these effects are negligible at flight scales. Completion of these experiments will advance the starlight suppression technology for starshades to TRL 5 and starshades will be ready for selection for the next exoplanet mission.

\appendix    


\section{Contrast Definition}
\label{apx:contrast_definition}
Contrast is defined as the amount of light within a resolution element of a telescope (at the image plane), divided by the peak brightness of the main light source as measured by that telescope when there is no starshade in place. The following equations define the contrast in terms of quantities measured in the lab; Table~\ref{tab:contrast_defn} provides descriptions of the variables used in the definition.

\begin{table}[ht]
\caption{List of variables used in contrast definition.}
\label{tab:contrast_defn}
\begin{center}
\begin{tabular}{c l}
\hline
\rule[-1ex]{0pt}{3.5ex} Symbol & Description\\
\hline\hline
\rule[-1ex]{0pt}{3.5ex} $C^i$ & Contrast at pixel $i$ \\
\rule[-1ex]{0pt}{3.5ex} $\Gamma^i$ & Transfer function mapping to pixel $i$\\
\rule[-1ex]{0pt}{3.5ex} $\gamma$ & Ratio of PSF peak after propagation through free space\\
	&\hspace{20pt} to that of the calibration mask\\
\rule[-1ex]{0pt}{3.5ex} $A$ & Peak value of apodization function $(=0.9)$\\
\rule[-1ex]{0pt}{3.5ex} $s^i$ & Counts collected in pixel $i$ during exposure of length $t$ [ct]\\
\rule[-1ex]{0pt}{3.5ex} $P$ & Laser power [W]\\
\rule[-1ex]{0pt}{3.5ex} $\nu$ & Neutral density filter transmission\\
\rule[-1ex]{0pt}{3.5ex} $\varepsilon$ & Transmission through atmosphere \\
\rule[-1ex]{0pt}{3.5ex} $\tau$ & Throughput of camera optics \\
\rule[-1ex]{0pt}{3.5ex} $Q$ & Quantum efficiency of detector [$e^-$/ph]\\
\rule[-1ex]{0pt}{3.5ex} $E$ & Photon energy [J/ph] \\
\rule[-1ex]{0pt}{3.5ex} $G$ & Camera inverse-gain [$e^-$/ct] \\
\rule[-1ex]{0pt}{3.5ex} $t$ & Exposure time [s]\\
\rule[-1ex]{0pt}{3.5ex} $m$ & Denotes {\it mask} measurement\\
\rule[-1ex]{0pt}{3.5ex} $u$ & Denotes {\it unocculted / calibration} measurement\\
\hline
\end{tabular}
\end{center}
\end{table}

Free space propagation is not possible in the confinements of the lab, thus we use a circular calibration mask to measure the unocculted brightness and convert to a free space brightness through modeling. In the following definitions, the subscript of a symbol will denote the observation mode, with $m$ denoting measurements made when the starshade {\it mask} is in place and $u$ denoting {\it unocculted} measurements when the calibration mask is in place.

We define the contrast at pixel $i$ as
\begin{equation}
    \begin{split}
	C^i &= \frac{\Gamma^i_m}{\Gamma^0_\text{free space}}\\
	&= \frac{\Gamma^i_m}{\gamma A^2\Gamma^0_u}\,,
	\end{split}
	\label{eq:contrast_0}
\end{equation}
which is a theoretical construct specifying the reduction in brightness the starshade mask provides, relative to the on-axis unocculted source.

The peak value of the apodization function ($A^2$) in the denominator accounts for the fraction of light that is blocked by the radial struts supporting the inner starshade. The value $\gamma$ is the ratio of the peak of the PSF after propagation through free space to that  after propagation through the circular calibration mask, and relates the contrast measured in the lab to that expected from a free-floating starshade. More details on this conversion can be found in Ref.~\citenum{Mile1a}.
The transfer function is tied to a measurement through the equation
\begin{equation}
	s^i_x = \left(\frac{P\nu\varepsilon Q\tau t \Gamma^i}{EG}\right)_x\,,
	\label{eq:counts}
\end{equation}
where $x\in \{m,u\}$. We drop the superscript on $s_u$ and assume it is on-axis. We assume there is no ND filter during starshade measurements ($\nu_m=1, \nu_u\equiv \nu$) and that the photon energy, camera gain, and camera throughput do not change between observation modes. Substituting Eq.~(\ref{eq:counts}) into Eq.~(\ref{eq:contrast_0}), we rewrite the contrast as
\begin{equation}
	C^i = \left(\frac{\nu s^i_mt_uP_u}{\gamma A^2s_ut_mP_m}\right)\left(\frac{\varepsilon_uQ_u}{\varepsilon_mQ_m}\right)\,.
	\label{eq:contrast_1}
\end{equation}

We assume that values in the right parentheses have the same mean between observation modes, but whose true value during a given observation is distributed normally around the mean with variance $\sigma^2$. In other words, $Q_u = Q_m\equiv Q$, $\sigma^2_{Q_u} = \sigma^2_{Q_m} \equiv\sigma^2_Q$, and similarly for $\varepsilon$. This simplifies the contrast definition to
\begin{equation}
	C^i = \frac{\nu s^i_mt_uP_u}{\gamma A^2s_ut_mP_m}\,.
	\label{eq:contrast_final}
\end{equation}

For model validation, we calculate the average contrast over $n$ pixels that lie in a photometric aperture of radius $\lambda/D$ centered at that pixel as
\begin{equation}
    \begin{split}
	C &= \frac{1}{n}\sum_i^nC^i\\
	&= \frac{1}{n} \frac{\nu t_uP_u}{\gamma A^2s_ut_mP_m}\sum_i^ns_m^i\\
	&\equiv \alpha\sum_i^ns_m^i\,,
	\end{split}
	\label{eq:average_contrast}
\end{equation}
where we have wrapped values independent of pixel into the constant variable $\alpha$. The uncertainty in $\alpha$ is given by
\begin{equation}
	 \frac{\sigma_{\alpha}^2}{\alpha^2}  = \frac{\sigma_{s_u}^2}{s_u^2} + \frac{\sigma_\gamma^2}{\gamma^2} + \frac{\sigma_\nu^2}{\nu^2} + \frac{\sigma_{P_u}^2}{P_u^2} + \frac{\sigma_{P_m}^2}{P_m^2} + 2\frac{\sigma_\varepsilon^2}{\varepsilon^2} + 2\frac{\sigma_Q^2}{Q^2}\,.
	\label{eq:constant_error}
\end{equation}

\new{The variance in the counts of the unocculted image, $\sigma_{s_u}^2$, is given by Eq.~(\ref{eq:cnt_var}). We estimate the values of the rest of the uncertainties of $\alpha$ in Ref.~\citenum{Mile1a} to find $\sigma_\alpha\sim2.5\%$. Assuming independent measurement errors, the uncertainty in the average contrast is propagated to
\begin{equation}
	\frac{\sigma_{C}^2}{C^2} = \frac{\sigma_{\alpha}^2}{\alpha^2} + \frac{\sum_i^n\sigma_{s_m^i}^2}{\left(\sum_i^ns^i_m\right)^2}\,,
	\label{eq:contrast_error}
\end{equation}
where $\sigma_{s_m^i}^2$ is the variance of pixel $i$ in the mask image.

The dominant contributions to the uncertainty in counts collected during each exposure $(s_u,s_m)$ are: photon noise from the source, background light, and detector dark noise and read noise. We can ignore noise from clock induced charge in the detector electronics, as this is estimated to be $<3\times10^{-3}$ events/pixel. Read noise is estimated from the variance of 2 bias frames of 10~$\upmu$s exposure time to be $\sigma_R$ = 3.20 e$^-$/pixel/frame. We combine the number of counts from background light and detector dark noise into the variable $d$, which is estimated from dark exposures taken with an equal exposure time. The uncertainty in the measurement of $s$ counts in a single image $j$ is given by
\begin{equation}
    \sigma_{s_j}^2 = \frac{s_j}{G} + 2\sigma_d^2 + \sigma_R^2 \,.
    \label{eq:cnt_var}
\end{equation}

For each observation mode, a number ($n_\mathrm{frames}$) of frames are taken and median-combined into a master image. The variance in $s$ counts obtained from $n_\mathrm{frames}$ frames is given by
\begin{equation}
    \sigma_s^2 = \frac{\sigma_{s_j}^2}{n_\mathrm{frames}}\,.
\end{equation}

Additional details on noise sources and calibrations can be found in Ref.~\citenum{Mile1a}.}

\section{The thick screen effect at flight scales}
\label{apx:skin_scaling}
\renewcommand*{\theHsection}{chY.\the\value{section}}

\new{In this appendix, we estimate how the thick screen effect scales with starshade size and argue the effect is negligible at flight scales. To start, we derive an expression for the change in intensity at the telescope due to the thick screen effect (assuming it induces a local change in amplitude only) and show that the expression is consistent with experimental results. We then apply the expression to the flight-scale starshade and show the effect is negligible.

\subsection{Derivation of intensity estimation}
We begin with the Fresnel-Kirchhoff diffraction equation and assume an incident plane wave with $z = Z_\mathrm{eff}$. The on-axis electric field $U$ is given by
\begin{equation}
    U = \frac{2\pi}{i\lambda z}\int_0^R A(r) \exp\left\{\frac{i\pi r^2}{\lambda z}\right\} r \, dr \,,
    \label{eq:apx_skin_1}
\end{equation}
where $A(r)$ is the circularly symmetric apodization function. We characterize the thick screen effect as a change in the apodization function ($\alpha$) relative to the nominal shape $(A_0)$ such that $A(r) = A_0(r) + \alpha(r)$. Equation~(\ref{eq:apx_skin_1}) then becomes
\begin{eqnarray}
    U &=& \frac{2\pi}{i\lambda z}\int_0^R \left[A_0(r) + \alpha(r)\right] \exp\left\{\frac{i\pi r^2}{\lambda z}\right\} r \, dr \nonumber \\
    &=& \frac{2\pi}{i\lambda z}\int_0^R A_0(r) \exp\left\{\frac{i\pi r^2}{\lambda z}\right\} r \, dr + \frac{2\pi}{i\lambda z}\int_0^R \alpha(r) \exp\left\{\frac{i\pi r^2}{\lambda z}\right\} r \, dr\nonumber \\
    &=& U_\mathrm{nominal} + \Delta U \,.
    \label{eq:apx_skin_2}
\end{eqnarray}

$U_\mathrm{nominal}$ is the nominal electric field in the scalar diffraction limit and $\Delta U$ is the change in electric field as a result of the thick screen effect. In Sec.~\ref{sec:thick_screen} we posited, and the models confirm\cite{Harness_2020}, that the presence of the thick screen induces a change in the electric field in a narrow ($\sim\lambda$) boundary layer around the edge. For now, we restrict ourselves to the case where the presence of the screen induces a change in amplitude only, resulting in a $\sim\lambda$ wide boundary layer around the starshade edge with zero transmission. Since the width of the boundary layer, which we define as $\delta$, is roughly constant for each edge (we neglect differences between polarization states), the change in the apodization function is related to the boundary layer width by
\begin{equation}
     \alpha(r) = \begin{cases}
        0, & r < R_0\\
        \dfrac{\delta N_p}{2\pi r}, & R_0 \le r \le R\,,\\
    \end{cases}
\end{equation}
where $N_p$ is the number of starshade petals, $R_0$ is the minimum radius at which the petals start, and we've acknowledged that there is no change in the apodization before the petals starts. The thick screen effect can now be written as
\begin{eqnarray}
    \Delta U &=& \frac{2\pi}{i\lambda z}\int_{R_0}^R \left(\frac{\delta N_p}{2\pi r}\right) \exp\left\{\frac{i\pi r^2}{\lambda z}\right\} r \, dr \nonumber \\
    &=& \frac{\delta N_p}{i\lambda z} \int_{R_0}^R  \exp\left\{\frac{i\pi r^2}{\lambda z}\right\} \, dr \,.
    \label{eq:apx_skin_3}
\end{eqnarray}

The integral in Eq.~(\ref{eq:apx_skin_3}) can readily be evaluated in terms of Fresnel integrals. We define the complex Fresnel integral as
\begin{equation}
    \mathcal{F}(u) \equiv C(u) + iS(u) \equiv \int_0^u e^{i\pi t^2/2} \, dt \,,
\end{equation}
and make the appropriate substitutions to write the thick screen effect as
\begin{equation}
        \Delta U = \left(\frac{\delta N_p}{i\lambda z}\right)\sqrt{\frac{\lambda z}{2}}\left[ \mathcal{F}\left(\sqrt{2N}\right) - \mathcal{F}\left(\sqrt{2N_0}\right)\right] \,,
\end{equation}
where $N$ is the Fresnel number at the starshade radius and $N_0$ is the Fresnel number at the start of the petals. The change in (on-axis) intensity due to the thick screen effect is then
\begin{equation}
    \left|\Delta U\right|^2 = \frac{\delta^2}{\lambda z}\frac{N_p^2}{2}\left| \mathcal{F}\left(\sqrt{2N}\right) - \mathcal{F}\left(\sqrt{2N_0}\right)\right|^2 \,.
    \label{eq:skin_contrast}
\end{equation}
We note that the leading factor looks like a Fresnel number across the width of the boundary layer.

\subsection{Comparison to experimental results}
We will now use Eq.~(\ref{eq:skin_contrast}) to estimate the change in intensity for the laboratory configuration and compare to experimental results of mask DW9. Equation~(\ref{eq:skin_contrast}) is the change in on-axis intensity at the telescope plane, so we will compare our estimates to the suppression values calculated from pupil plane images. Suppression is the total intensity incident on the telescope's aperture when the starshade is occulting the star, relative to the total intensity of the unblocked star.

At a wavelength of $\lambda=638$~nm, finite-difference time-domain (FDTD) simulations of the mask DW9 geometry yield an average change in amplitude consistent with a boundary layer width of $\delta=0.45$\micron (averaging over polarization states). We input the parameters from Table~\ref{tab:apod_design} into Eq.~(\ref{eq:skin_contrast}) to estimate the change in intensity to be $\left|\Delta U\right|^2 = 1.6\times10^{-8}$. Figure~\ref{fig:suppression} shows the suppression plot (image of the pupil plane) for DW9 at $\lambda=638$~nm. These data were taken without any polarizing elements, so represent a rough average over polarization states. The peak suppression is $1.5\times10^{-7}$ and the average over the aperture is $6.3\times10^{-8}$, which is greater than, but within an order of magnitude of, that predicted by Eq.~(\ref{eq:skin_contrast}). These results show our estimate of the thick screen effect is within reason and we note that Eq.~(\ref{eq:skin_contrast}) was a lower limit as it assumed an amplitude-only change in field due to the thick screen.
\begin{figure}[htb]
\begin{center}
\begin{tabular}{c}
  \includegraphics[width=0.75\linewidth]{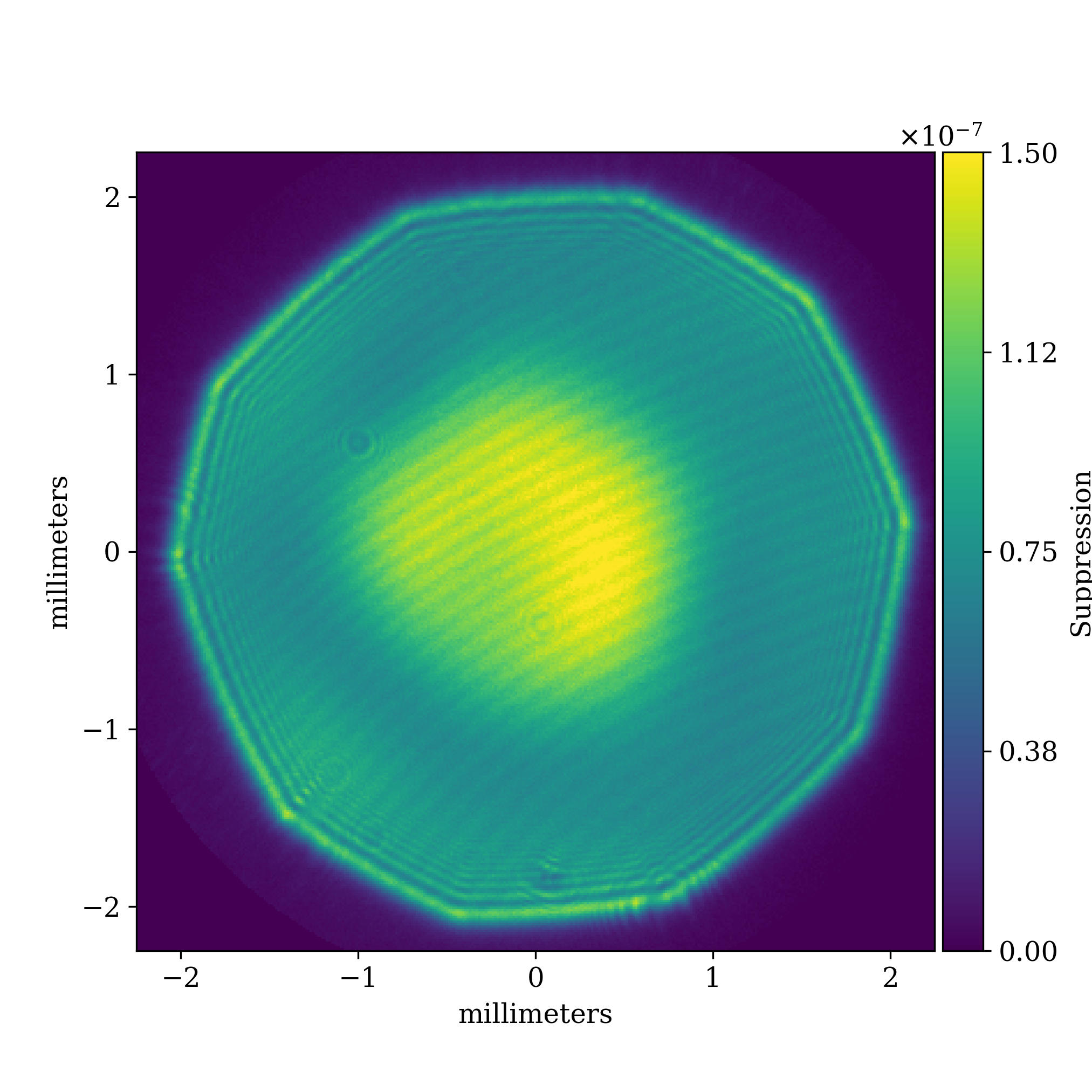}
\end{tabular}
\end{center}
\caption
{ \label{fig:suppression}
  Suppression (pupil plane) image of DW9 at $\lambda=638$~nm, without any polarizing elements.}
\end{figure}

\subsection{Scaling up to flight}
By examining Eq.~(\ref{eq:skin_contrast}), we can see why the non-scalar diffraction should be negligible at flight scales. Scaling from the lab configuration to that of flight, the Fresnel numbers are the same and $\delta$ will be roughly the same (the optical edges of the flight design are of similar thickness to those in the lab), so the thick screen effect intensity goes as $\left|\Delta U\right|^2\sim z^{-1}$. The effective starshade-telescope separation for flight is 10$^6\times$ that in the lab (the starshades are 1000$^\mathrm{th}$ scale and the separation scales quadratically with size for a given Fresnel number), so we can expect the non-scalar diffraction to be 10$^6\times$ lower in intensity.

Replacing $\lambda z$ in the denominator of Eq.~(\ref{eq:skin_contrast}) with $R^2N$ gives $\left|\Delta U\right|^2\sim \delta^2 / R^2$, which is the same argument given in Sec.~\ref{sec:thick_screen} that the thick screen effect goes as the area of the boundary layer relative to the transmission area. A final observation: the leading factor in Eq.~(\ref{eq:skin_contrast}) is similar to a Fresnel number across the boundary layer, which becomes increasingly small at large separations (because the width remains constant), meaning there is little phase variance across the width.}

\section{Optical Model Description}
\label{apx:model_description}

The optical model summarized here is presented in full in Ref.~\citenum{Harness_2020}. The scalar diffraction problem we are solving uses the standard Fresnel and Kirchhoff assumptions that allow us to express the diffraction propagation as a Fourier transform. Because the inner starshade is connected to the outer supporting wafer via the radial struts, the starshade pattern can be treated as individual apertures and standard Fourier techniques sufficiently resolve the starshade\cite{Harness_2018}. If the coordinates of the screen and observation planes are given as $\left(\xi,\eta\right)$ and $\left(x,y\right)$, respectively, we can write the Fresnel-Kirchhoff diffraction equation as
\begin{equation}
    U\left(x,y\right) \propto \mathcal{F}\left[U_0\left(\xi,\eta\right)\cdot A\left(\xi,\eta\right)\cdot e^{\frac{ik}{2z}\left(\xi^2+\eta^2\right)}\right]_{\left(x,y\right)} \,,
    \label{eq:FK_diffraction}
\end{equation}
where $U_0$ is the initial field incident on the screen, $A$ represents the aperture function of the screen, and $\mathcal{F}\left[\cdot\right]_{\left(x,y\right)}$ is the Fourier transform. Using the Kirchhoff boundary values, $A$ is a binary function that is 0 on the screen and 1 in the aperture. Our method implements non-scalar diffraction into this model by replacing the boundary values of $A$ with a complex field that arises from local diffraction at the edge of the screen. The presence of the screen only affects its immediate surrounding, so we only change the values of $A$ in a narrow ($\sim10\lambda$ wide) seam around the edge of the screen, similar to the method proposed by Braunbek\cite{Braunbek}.

\begin{figure}[htb]
\begin{center}
\begin{tabular}{c}
  \includegraphics[width=0.9\linewidth]{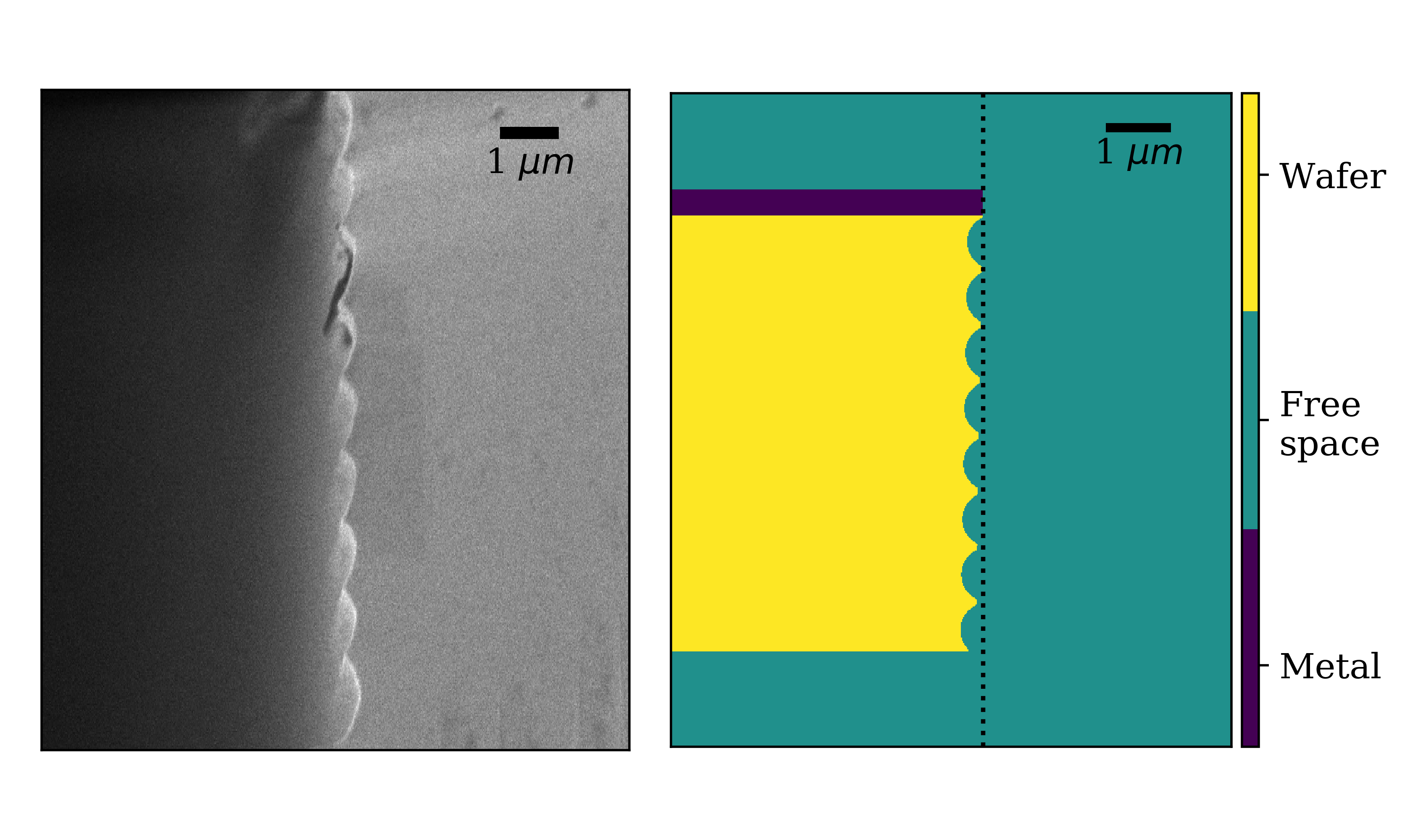}
\end{tabular}
\end{center}
\caption
{ \label{fig:sidewalls}
  Left: SEM image of the scalloped vertical profile of the wafer edge for a manufactured mask. Right: electric permittivity map of the \emph{Meep}\cite{Meep} FDTD simulation cell displaying the material geometry.}
\end{figure}

The field in the seam around the edge is solved for via an FDTD simulation\cite{Meep} of light propagating past the edge of a metal-coated silicon wafer. This simulation solves Maxwell's equations near the edge and allows us to simulate the geometry and material properties of the manufactured starshade masks. In order to get the model to match the results of Sec.~\ref{sec:polarization}, we had to account for scalloping and tapering of the mask sidewalls left by the etching process. The edge properties that produced the best fit to the data (1$^\circ$ taper angle and scallops 0.8\micron tall and 0.2\micron deep) were confirmed to be those of the real mask from scanning electron microscope images of the edge of a manufactured mask. Figure~\ref{fig:sidewalls} shows the simulated geometry of the edge along with SEM images of a manufactured mask, confirming we are properly simulating the complex edge structures.


\subsection*{Disclosures}
Some of the work presented here is included in the S5 Milestone Reports of Refs.~\citenum{Mile1a, Mile1b} and in Ref.~\citenum{Harness_2019}. Anthony Harness is a guest editor for this special section.

\acknowledgments
The authors thank Jonathan Arenberg for his helpful and constructive comments. This work was performed in part at the Jet Propulsion Laboratory, California Institute of Technology under a contract with the National Aeronautics and Space Administration. Starshade masks were manufactured using the facilities at the Microdevices Lab at JPL. This project made use of the resources from the Princeton Institute for Computational Science and Engineering (PICSciE) and the Office of Information Technology's High Performance Computing Center and Visualization Laboratory at Princeton University.


\bibliography{main}   
\bibliographystyle{spiejour}   

\newpage
\vspace{2ex}\noindent\textbf{Anthony Harness} is an Associate Research Scholar in the Mechanical and Aerospace Engineering Department at Princeton University. He received his Ph.D. in Astrophysics in 2016 from the University of Colorado Boulder. He currently leads the experiments at Princeton validating starshade optical technologies.

\vspace{2ex}\noindent\textbf{Stuart Shaklan} is the supervisor of the High Contrast Imaging Group in the Optics Section of the Jet Propulsion Laboratory. He received his Ph.D. in Optics at the University of Arizona in 1989 and has been with JPL since 1991.

\vspace{2ex}\noindent\textbf{Phil Willems} is an Optical Engineer at the Jet Propulsion Laboratory, where he is the manager of the S5 Starshade Technology Development Activity. He received his BS degree in physics from the University of Wisconsin-Madison in 1988, and his PhD degree in physics from the California Institute of Technology in 1997.

\vspace{2ex}\noindent\textbf{N. Jeremy Kasdin} is the Assistant Dean for Engineering at the University of San Francisco and the Eugene Higgins professor of Mechanical and Aerospace Engineering, emeritus, at Princeton University. He received his Ph.D. in 1991 from Stanford University. After being the chief systems engineer for NASA's Gravity Probe B spacecraft, he joined the Princeton faculty in 1999 where he researched high-contrast imaging technology for exoplanet imaging. From 2014 to 2016 he was Vice Dean of the School of Engineering and Applied Science at Princeton. He is currently the Adjutant Scientist for the coronagraph instrument on NASA’s Wide Field Infrared Survey Telescope.


\listoffigures
\listoftables

\end{spacing}
\end{document}